\documentclass[aps, prd, twocolumn, amsmath, floats,floatfix, superscriptaddress, nofootinbib]{revtex4}
\usepackage{graphicx}
\usepackage{bm}
\usepackage{hyperref}
\usepackage[utf8]{inputenc}
\hypersetup{
  colorlinks=true,        
  linkcolor=blue,         
  citecolor=cyan,         
  urlcolor=cyan            
}
\usepackage{amsmath,amsfonts,amssymb}
\usepackage{times}
\usepackage[T1]{fontenc}
\usepackage{float}
\usepackage[normalem]{ulem}
\usepackage{multirow}
\usepackage{soul}

\newcommand{\pr}[1]{\ensuremath{\left[#1\right]}}
\newcommand{\pc}[1]{\ensuremath{\left(#1\right)}}
\newcommand{\chav}[1]{\ensuremath{\left\{#1\right\}}}

\DeclareMathOperator*{\argmin}{arg\,min} 
\DeclareMathOperator*{\argmax}{arg\,max}

\newcommand{\be}{\begin{equation}}
\newcommand{\ee}{\end{equation}}
\newcommand{\bea}{\begin{eqnarray}}
\newcommand{\eea}{\end{eqnarray}}

\DeclareMathOperator{\E}{\mathbb{E}}

\makeatletter
\newcommand*{\rom}[1]{\expandafter\@slowromancap\romannumeral #1@}
\makeatother

\begin{document}
\title{Decoding Neutron Star Observations: Revealing Composition through Bayesian Neural Networks}

\author{Valéria Carvalho}
\email{val.mar.dinis@gmail.com}
\affiliation{CFisUC, 
	Department of Physics, University of Coimbra, P-3004 - 516  Coimbra, Portugal}

\author{Márcio Ferreira}
\email{marcio.ferreira@uc.pt}
\affiliation{CFisUC, 
	Department of Physics, University of Coimbra, P-3004 - 516  Coimbra, Portugal}
 
\author{Tuhin Malik}
\email{tm@uc.pt}
\affiliation{CFisUC, 
	Department of Physics, University of Coimbra, P-3004 - 516  Coimbra, Portugal}
	
\author{Constança Providência}
\email{cp@uc.pt}
\affiliation{CFisUC, 
	Department of Physics, University of Coimbra, P-3004 - 516  Coimbra, Portugal}

\date{\today}

\begin{abstract} 
We exploit the great potential offered by Bayesian Neural Networks (BNNs) to directly decipher the internal composition of neutron stars (NSs) based on their macroscopic properties. By analyzing a set of simulated observations, namely NS radius and tidal deformability, we leverage BNNs as effective tools for inferring the proton fraction and sound speed within NS interiors. To achieve this, several BNNs models were developed upon a dataset of $\sim$ 25K nuclear EoS within a relativistic mean-field framework, obtained through Bayesian inference that adheres to minimal low-density constraints. Unlike conventional neural networks, BNNs possess an exceptional quality: they provide a prediction uncertainty measure. To simulate the inherent imperfections present in real-world observations, we have generated four distinct training and testing datasets that replicate specific observational uncertainties. Our initial results demonstrate that BNNs successfully recover the composition with reasonable levels of uncertainty. Furthermore, using mock data prepared with the DD2, a different class of relativistic mean-field model utilized during training, the BNN model effectively retrieves the proton fraction and speed of sound for neutron star matter.

\end{abstract}

\maketitle
\definecolor{vc}{rgb}{0.0, 0.5, 0.69}
\section{Introduction}

The extreme matter conditions inside neutron stars (NSs) are impossible to recreate in terrestrial laboratories, making the   
equation of state (EoS) of dense and asymmetric nuclear matter (realized inside NSs) an interesting and still unknown quantity. 
Modeling NS matter is restricted by constraints coming from the observations of massive NSs:  PSR~J1614-2230   \cite{Demorest2010,Fonseca2016,Arzoumanian2017} with  $M = 1.908 \pm~ 0.016 M_{\odot}$, PSR~ J0348 - 0432 with $M = 2.01 \pm~ 0.04~ M_{\odot}$ \cite{Antoniadis2013},  PSR J0740+6620 with $M = 2.08 \pm~ 0.07~ M_{\odot}$  \cite{Fonseca:2021wxt}, and J1810+1714 with $M = 2.13 \pm~ 0.04~ M_{\odot}$  \cite{Romani:2021xmb}. Additionally, theoretical calculations, such as chiral effective field theory (cEFT), are applicable only at very low densities, while perturbative quantum chromodynamics (pQCD) is reliable for extremely high densities. Recently, multi-messenger astrophysics has become an exciting field, allowing  to understand NSs physics by connecting information carried by different sources, such as gravitational waves, photons, and neutrinos.  The detection by LIGO/Virgo collaboration of compact binary coalescence events, such as GW170817 \cite{Abbott:2018wiz} and GW190425 \cite{Abbott:2020khf}, allowed us to further constrain the EoS of NS matter. Recent results from NICER (Neutron star Interior Composition ExploreR) on the PSR J0030+045 pulsar mass \cite{Riley_2019,Miller19} and PSR J0740+6620 radius \cite{Riley2021,Miller2021,Raaijmakers2021} were relevant in restricting the possible neutron stars physics. Future expected observations from experiments such as the enhanced X-ray Timing and Polarimetry mission (eXTP) \cite{eXTP,eXTP:2018anb}, the (STROBE-X) \cite{STROBE-X}, and Square Kilometer Array \citep{SKA} telescope will allow for the determination of NSs radii and masses with a few \% uncertainty.\\

Numerous statistical methods have been extensively explored to determine the most probable EoS based on observational data of NSs. These methods include Bayesian inference \cite{Malik:2022zol,Malik:2023mnx} and Gaussian processes \cite{Essick:2020flb}. However, even if the EoS is known with high precision, the challenge remains in constraining the composition of neutron star matter. Previous studies have highlighted the impossibility of recovering nuclear matter properties solely from the $\beta$-equilibrium EoS without knowledge of the compositions (or symmetry energy at high densities) \cite{Tovar2021,Imam2021,Mondal2021} or without information about the EoS of symmetric nuclear matter in conjunction with compositions  \cite{Essick2021}. However, these studies were either limited to meta-models or based on restricted models. Motivated by this, we aim to construct an artificial neural network that directly maps NS observational properties to EoS composition using a large set of EoS derived from the Relativistic Mean Field (RMF) approach. \\

Deep learning is another field that has become a buzzword in trying to solve the dense matter EoS problems,  and all kinds of physics problems \cite{fujimoto2021extensive,Fujimoto_2018,Fujimoto_2020,soma2022reconstructing,Soma_2022,chatterjee2023analyzing,morawski2020neural,PhysRevC.106.065802,Krastev_2022,han2021bayesian,Han:2022sxt,krastev2023deep,Traversi_2020,ferreira2021unveiling,ferreira2022extracting,gonccalves2022machine,farrell2023deducing,Thete:2022eif,murarka2022neutron,Zhou:2023cfs}.
The inference problem of determining the EoS from observational data can be roughly divided into two main categories: the reconstructing of the EoS, pressure or speed of sound, from either mass radius or tidal deformability, \cite{Fujimoto_2018,Fujimoto_2020,fujimoto2021extensive,chatterjee2023analyzing,morawski2020neural,PhysRevC.106.065802,soma2022reconstructing,Soma_2022,han2021bayesian,Han:2022sxt,Traversi_2020,gonccalves2022machine,farrell2023deducing,Zhou:2023cfs}, or focus directly on the nuclear matter saturation properties \cite{ferreira2021unveiling,ferreira2022extracting,krastev2023deep,Krastev_2022,Thete:2022eif,murarka2022neutron}. As  an example, Fujimoto et. al. \cite{Fujimoto_2018,Fujimoto_2020,fujimoto2021extensive} explored a framework based on neural networks (NNs) where  
an observational set of NS mass, radius and respective variances was used as input and the speed of sound squared as  output. A similar perspective was followed in \cite{chatterjee2023analyzing}. The works \cite{krastev2023deep,Krastev_2022} fall into the second category: determining specific saturation properties of nuclear matter, in this case, the density dependence of the nuclear symmetry energy, directly from observational NS data.\\

However, the majority of these NNs based models face a considerable drawback, namely, the lack of uncertainty quantification. Questions such as \textit{how confident is a model about its predictions?} is  the main focus  of the present work, in which the uncertainty modeling is explored 
by implementing a very appealing approach called Bayesian Neural Networks (BNNs), \cite{mackay1992practical}. BNNs have already started being used in different fields of physics \cite{bollweg2020deep,lin2021detection}, and have been shown useful in uncertainty quantification.  
Our goal is to implement an inference framework that gives a prediction uncertainty to any model prediction. We analyze the density dependence of the proton fraction and speed of sound inside NS matter. For that, several synthetic datasets made of "mock" observational data will be constructed and the impact of adding information on the tidal deformability into the model predictions will be analyzed. Let us make clear an important distinction between our work and the majority of studies: instead of applying a widely used approach of parameterizing the EoS, e.g., with polytropes, we used a specific family of nuclear models to construct a set of possible EoS.  Despite their flexibility in exploring the entire region of possible EoS, the generic agnostic parametrizations of the EoS are unable to model and track the different degrees of freedom inside NS. 
The use of a microscopic model has the crucial advantage of accessing the density dependence of all degrees of freedom and thus unable us to analyze the proton fraction.  \\

The paper is organized as follows. A basic introduction to BNNs is presented in Sec \ref{bnn}. The family of nuclear models chosen is presented in Sec. \ref{models} and also the Bayesian inference framework employed to construct the EoS dataset. The generation of the synthetic observation datasets is explained in Sec. \ref{dataset}. The model results for the proton fraction and speed of sound are discussed in Sec. \ref{results}, and lastly, the conclusions are drawn in Sec. \ref{conclusions}.

\section{Bayesian Neural Networks \label{bnn}}

Despite the great success of (feedforward) neural networks (NNs) in different fields of science, they come with some drawbacks that require special attention. NNs are susceptible to over-fitting and are unable to access the uncertainty of its predictions, which may lead to overconfident predictions. 
Bayesian Neural Networks (BNNs) is a Bayesian approach framework 
that introduces stochastic weights to NNs making them 
uncertainty-aware models \cite{jospin2022hands}.\\

NNs are capable of representing arbitrary functions and are composed, in their simplest architecture, of a sequence of blocks where a linear transformation is followed by a nonlinear operation (activation functions). To simplify the notation, let us denote a NN by $\bm{y}=f_{\bm{\theta}}(\bm{x})$, where $\bm{\theta}=(\bm{W},\bm{b})$ represent all NN weights. The vectors $\bm{W}$ and $\bm{b}$ denote, respectively, the connections (weights) and bias of all linear transformations of the network, which define completely the NN model. 
Training the NN consists in determining the numerical procedure (back-propagation algorithm) of finding the $\bm{\theta}^{*}$ that minimizes a chosen cost function on the training data. This traditional approach of estimating a single model  defined by $\bm{\theta}^{*}$ ignores all other possible parametrizations $\bm{\theta}$.\\

BNNs simulate multiple possible NNs models by introducing stochastic weights. These networks operate by first choosing a functional model, i.e., a network architecture, and then the stochastic model, i.e., the probability distributions for the weights. Bayesian inference is then required to train the network by defining the likelihood function of the observed data, $P(D|\bm{\theta})$, and the prior probability distribution over the model parameters, $P(\bm{\theta})$. It is then possible to employ Bayes theorem and obtain the posterior probability distribution, i.e. the probability of the model parameters given the data: 
\begin{equation}\label{eq:bayes}
 P(\bm{\theta}|D) = \frac{P(D|\bm{\theta})P(\bm{\theta})}{ P(D) } 
\end{equation}
where $P(D)=\int_{\bm{\theta}'}{P(D|\bm{\theta}')P(\bm{\theta}')d\bm{\theta}'}$  is the evidence.
Having a distribution on the weights, the BNNs predictions become a  Bayesian model average: the probability distribution of some unknown  $\bm{y^*}$ given an input $\bm{x^*}$ is
\begin{equation} \label{eq:baye_pred}
    P(\bm{y^*}|\bm{x^*}, D) = \int_{\bm{\theta}} P(\bm{y^*}|\bm{x^*},\bm{\theta})P(\bm{\theta}|D) d \bm{\theta}.
\end{equation}
$P(\bm{y^*}|\bm{x^*},\bm{\theta})$ is considered to be the likelihood of our data, the distribution that comes out of the network and captures the noise present in our data, and $P(\bm{\theta}|D)$ is the posterior distribution of our weights, that brings up the uncertainty on the model.
Another advantage of using these networks is that they capture two types of uncertainty, aleatoric uncertainty, uncertainty on the data, and epistemic uncertainty, uncertainty on the model estimation defined as $P(\bm{y^*}|\bm{x^*},\bm{\theta})$ and $P(\bm{\theta}|D)$ respectively.
However, solving Eq. \ref{eq:baye_pred} is a very complex task  because the posterior $P(\bm{\theta}|D)$ depends on the evidence $P(D)=\int_{\bm{\theta}'}{P(D|\bm{\theta}')P(\bm{\theta}')d\bm{\theta}'}$, which is a non-analytic expression that requires marginalizing over all model parameters. 
Fortunately, there are multiple ways of tracking it by using either Markov Chain Monte Carlo or variational inference (more information can be found in \cite{jospin2022hands}).
We are going to implement the variational inference method that is presented in the following. 

\subsection{Varitional inference formalism\label{sec:vi}}

The variational inference method aims to approximate a variational posterior $q_{\bm{\phi}}(\bm{\theta})$ to the real posterior $P(\bm{\theta}|D)$ by using the Kullback-Leibler (KL) divergence. The KL divergence is a measure of dissimilarity between two probability distributions. It approaches zero when the variational posterior and the true posterior are identical and is positive otherwise. Fundamentally, we want to find the variational posterior that corresponds to the minimum value of the KL divergence between the variational posterior and the true posterior:
\begin{equation}
    q_{\phi^*}=\argmin_{q_\phi} \text{ KL}(q_\phi(\bm{\theta})||P(\bm{\theta}|D)), 
\end{equation}
where the KL divergence is defined by
\begin{align}
   \text{KL}(q_\phi(\bm{\theta})||P(\bm{\theta}|D))&= \E_{q_\phi(\bm{\theta})}\pr{\log \left( \frac{q_\phi(\bm{\theta})}{P(\bm{\theta}|D)}\right)} \\
   &= \int_{\bm{\theta}} q_\phi(\bm{\theta}) \log \left( \frac{q_\phi(\bm{\theta})}{P(\bm{\theta}|D)}\right) d\bm{\theta}
\end{align}
In order for the dependence on the true posterior to disappear, the last equation can be rewritten, with the help of Bayes rule Eq. \ref{eq:bayes}, as  
\begin{align}
\text{KL}&(q_\phi(\bm{\theta})||P(\bm{\theta}|D))= \int_{\bm{\theta}} q_\phi(\bm{\theta}) \log \left( \frac{q_\phi(\bm{\theta}) P(D)}{P(D|\bm{\theta}) P(\bm{\theta})}\right) d\bm{\theta} \nonumber\\
  &=\text{KL}(q_\phi(\bm{\theta})||P(\bm{\theta})) 
   - \E_{q_\phi(\bm{\theta})}(\log P(D|\bm{\theta}))+ \log P(D) \nonumber\\
     &= F(D,\phi) +\log P(D), 
\end{align}
where $F(D,\phi)=\text{KL}(q_\phi(\bm{\theta})||P(\bm{\theta})) - \E_{q_\phi(\bm{\theta})}(\log P(D|\bm{\theta}))$ is called the variational free energy. We end up with
\begin{equation}
    \text{KL}(q_\phi(\bm{\theta})||P(\bm{\theta}|D))=F(D,\phi) +\log P(D).
    \label{eq:eqq}
\end{equation}
As the last term, $\log P(D)$, does not depend on the variational posterior and his parameters, which we are optimizing, 
minimizing $\text{KL}(q_\phi(\bm{\theta})||P(\bm{\theta}|D))$ with respect to $\phi$ is the same as minimizing $F(D,\phi)$. ELBO is another important quantity, which stands for evidence lower bound, and it is defined as the negative free energy, i.e., $\text{ELBO}=-F(D,\phi)$. Equation \ref{eq:eqq} can then be rewritten as
\begin{equation}
\text{KL}(q_\phi(\bm{\theta})||P(\bm{\theta}|D))=-\text{ELBO} +\log P(D),
\end{equation}
or
\begin{equation}
\text{ELBO}= -\text{KL}(q_\phi(\bm{\theta})||P(\bm{\theta}|D)) +\log P(D).
\end{equation}
ELBO is called the lower bound of the evidence because ELBO $\leq \log P(D)$.
In other words, $\text{KL}(q_\phi(\bm{\theta})||P(\bm{\theta}|D))$ is minimized by maximizing the evidence lower bound.  \\\\
In the end, our optimization objective resumes to 
\begin{align*}
q_{\phi^*}&=\argmin_{q_\phi} \text{ KL}(q_\phi(\bm{\theta})||P(\bm{\theta}|D))\\
&=\argmax_{q_{\phi}}\text{ ELBO} =\argmin_{q_{\phi}} F(D,\phi)\\
&=\argmin_{q_{\phi}}\pr{\text{KL}(q_\phi(\bm{\theta})||P(\bm{\theta})) - \E_{q_\phi(\bm{\theta})}(\log P(D|\bm{\theta}))}.
\end{align*}

The above general formalism is applied to our specific case, where we have chosen a multivariate Gaussian for the variational posterior, $q_\phi(\bm{\theta})=\mathcal{N}(\bm{\mu}_q,\bm{\Sigma}_q)$, and a multivariate Gaussian with diagonal covariance matrix for the prior, $P(\bm{\theta})=\mathcal{N}(\mathbf{0},\mathbf{I})$ \footnote{ The choice of different probabilistic distributions and its impact on the inference results is left for future studies.}. The final loss function we are trying to minimize uses Monte Carlo sampling to obtain the expected values, where $\bm{\theta}^{(n)}$ is being sampled from the variational posterior, $q_\phi(\bm{\theta})$, and for our specific model,  also the exact value of the KL divergence between two Gaussian's with covariance matrix is used, 
\begin{align}
    F(D,\phi)&=\frac{1}{2D_s}\left[\left( -\log 
    \det(\bm{\Sigma}_q) \right)- k + \operatorname{tr}\left(\bm{\Sigma}_q\right) + (\bm{\mu}_q)^T(\bm{\mu}_q)\right]\nonumber\\
    &- \frac{1}{B}\sum_{i=1}^B\frac{1}{N}\sum_{n=1}^N \log P(\bm{y_i}|\bm{x_i},\bm{\theta}^{(n)}),\label{eq:final_loss}
\end{align}
where $B$ is the number of points of the mini-batch, $D_s$ is the number of points of the training dataset, $k$ is the dimension of the identity matrix of the prior and $N$ is the number of samples we take of the variational posterior (we used $N=10^4$). One aspect that stands out in these networks is how back-propagation works. Without going into much detail, the back-propagation updates $\bm{\mu}_q$ and $\bm{\Sigma}_q$ for each of the networks's parameters (more information can be found in \cite{blundell2015weight}).
Once the network is trained and the best mean $\bm{\mu}_q$ and covariance matrix $\bm{\Sigma}_q$ are obtained for the distribution of the parameters, Eq. \ref{eq:baye_pred} becomes solvable and predictions are obtained using Monte Carlo estimations
\begin{align}
P(\bm{y^*}|\bm{x^*}, D) = &\int_{\bm{\theta}} P(\bm{y^*}|\bm{x^*},\bm{\theta})q_\phi(\bm{\theta}) d \bm{\theta} \\
=& \frac{1}{N} \sum_{n=1}^{N} P(\bm{y^*}|\bm{x^*},\bm{\theta}^{(n)}), \quad \bm{\theta}^{(n)} \sim q_\phi(\bm{\theta}).
\end{align}
The mean $\hat{\bm{\mu}}$ and variance $\hat{\bm{\sigma}}^2$ vectors of the predicting distribution $P(\bm{y^*}|\bm{x^*}, D)$ can be calculated, for a fixed $\bm{x^*}$, by applying the law of total expectation and total variance. From
\begin{equation}
    \E [\bm{y^*}|\bm{x^*},D]= \E_{q_\phi (\bm{\theta})}\left [\E[\bm{y^*}|\bm{x^*},\bm{\theta}]\right]
\end{equation}
and
\begin{align}
    {\mathrm{Var}}\,[\bm{y^*}|\bm{x^*},D] =& \E_{q_\phi (\bm{\theta})} \left [{\mathrm{Var}}\,[\bm{y^*}|\bm{x^*},\bm{\theta}]\right]\nonumber\\
    &- {\mathrm{Var}}\,_{q_\phi (\bm{\theta})} \left[ \E[\bm{y^*}|\bm{x^*},\bm{\theta}] \right],
\end{align}

we obtain 
\begin{equation}
    \hat{\bm{\mu}}=\frac{1}{N}\sum_{n=1}^N\hat{\bm{\mu}}_{\bm{\theta}_n},
\end{equation}
and
\begin{equation}
    \hat{\bm{\sigma}}^2= \underbrace{\frac{1}{N}\sum_{n=1}^N\hat{\bm{\sigma}}_{\bm{\theta}_n}^2}_{\text{Aleatoric uncertainty}}+\underbrace{\frac{1}{N}\sum_{n=1}^N(\hat{\bm{\mu}}_{\bm{\theta}_n} - \hat{\bm{\mu}})\odot(\hat{\bm{\mu}}_{\bm{\theta}_n} - \hat{\bm{\mu}})}_{\text{Epistemic uncertainty}},
    \label{eq:errors}
\end{equation}
where $\odot$ denotes element-wise multiplication.
The predicted variance captures both epistemic and aleatoric uncertainties \cite{olivier2021bayesian}.

\section{Nuclear models \label{models}}
A field theoretical approach is adopted to calculate a dataset of nuclear equations of state (EoSs). The approach incorporates self-interactions and mixed meson terms within a relativistic mean field (RMF) description. A wide and reasonable region of the parameter space is considered, providing an accurate representation of presently known nuclear properties. Including non-linear terms is crucial for determining the density dependence of the EoS. In this treatment, the nucleons interact through the exchange of scalar-isoscalar mesons ($\sigma$), vector-isoscalar mesons ($\omega$), and vector-isovector mesons ($\rho$). The Lagrangian governing the baryonic degrees of freedom can be expressed as follows:
$
  \mathcal{L}=   \mathcal{L}_N+ \mathcal{L}_M+ \mathcal{L}_{NL}
$
with
\begin{align*}
    \mathcal{L}_{N}=& \bar{\Psi}\Big[\gamma^{\bm{\mu}}\left(i \partial_{\bm{\mu}}-g_{\omega} \omega_{\bm{\mu}}-
 g_{\varrho} {\boldsymbol{t}} \cdot \boldsymbol{\varrho}_{\bm{\mu}}\right) \\
&-\left(m-g_{\sigma} \sigma\right)\Big] \Psi \\
\mathcal{L}_{M}=& \frac{1}{2}\left[\partial_{\bm{\mu}} \sigma \partial^{\bm{\mu}} \sigma-m_{\sigma}^{2} \sigma^{2} \right] \\
&-\frac{1}{4} F_{\bm{\mu} \nu}^{(\omega)} F^{(\omega) \bm{\mu} \nu} 
+\frac{1}{2}m_{\omega}^{2} \omega_{\bm{\mu}} \omega^{\bm{\mu}} \nonumber\\
&-\frac{1}{4} \boldsymbol{F}_{\bm{\mu} \nu}^{(\varrho)} \cdot \boldsymbol{F}^{(\varrho) \bm{\mu} \nu} 
+ \frac{1}{2} m_{\varrho}^{2} \boldsymbol{\varrho}_{\bm{\mu}} \cdot \boldsymbol{\varrho}^{\bm{\mu}}\\
    			\mathcal{L}_{NL}=&-\frac{1}{3} b ~m~ g_\sigma^3 (\sigma)^{3}-\frac{1}{4} c g_\sigma^4 (\sigma)^{4}+\frac{\xi}{4!} g_{\omega}^4(\omega_{\bm{\mu}}\omega^{\bm{\mu}})^{2} \nonumber\\&+\Lambda_{\omega}g_{\varrho}^{2}\boldsymbol{\varrho}_{\bm{\mu}} \cdot \boldsymbol{\varrho}^{\bm{\mu}} g_{\omega}^{2}\omega_{\bm{\mu}}\omega^{\bm{\mu}},
\end{align*}
where the field $\Psi$ denotes the Dirac spinor that describes the nucleon doublet (neutron and proton) with a  bare mass $m$,  $\gamma^{\bm{\mu}} $ are the Dirac matrices, and $\boldsymbol{t}$ is the isospin operator. The vector meson  tensors are defined as   $F^{(\omega, \varrho)\bm{\mu} \nu} = \partial^{ \bm{\mu}} A^{(\omega, \varrho)\nu} -\partial^ \nu A^{(\omega, \varrho) \bm{\mu}}$.  
The couplings of the nucleons to the meson fields $\sigma$, $\omega$, and $\varrho$ are denoted by $g_{\sigma}$, $g_{\omega}$, and $g_{\varrho}$, respectively. The meson masses  are given by  $m_\sigma$, $m_\omega$, and $m_\varrho$. More information on the specifics of the model can be found in \cite{Malik:2023mnx} and references therein. 
The parameters $g_\sigma$, $g_\omega$, $g_\rho$, $b$, $c$, $\xi$, and $\Lambda_{\omega}$ are systematically sampled within a Bayesian framework, adhering to minimal constraints imposed by several nuclear saturation properties. Furthermore, these parameters are subject to the conditions of the neutron star maximum mass exceeding 2$M_\odot$, as well as the EoS for low-density pure neutron matter, which is meticulously generated through a precise N$^3$LO calculation in chiral effective field theory. A detailed discussion on these aspects will be presented in the subsequent subsection.

\subsection{The Bayesian setup}
\label{sec:bayesian_setup}
Based on observed or fitted data, a prior belief (expressed as a prior distribution) is updated using Bayesian inference. The posterior distribution is derived according to Bayes' theorem \cite{Gelman2013}. In order to establish a Bayesian parameter optimization system, four key components must be defined: the prior, the likelihood function, the fit data, and the sampler.

{\it The Prior:-} A broad range of nuclear matter saturation properties is carefully considered in the prior domain of the adopted RMF model. As a result of Latin hypercube sampling, we determine the prior range in our Bayesian setup. Uniform priors are chosen for each parameter, as described in Table \ref{tab2}.

\begin{table}[!ht]
\caption{We use a uniform prior range for the parameters of the RMF models. Specifically, B and C are $b \times 10^3$ and $c \times 10^3$, respectively. Distribution minimums and maximums are indicated by 'min' and 'max' respectively.}
\label{tab2}
\setlength{\tabcolsep}{15.5pt}
\renewcommand{\arraystretch}{1.1}
\begin{tabular}{cccc}
\toprule
\multirow{2}{*}{No} & \multirow{2}{*}{Parameters}  & \multicolumn{2}{c}{\it Set 0} \\ \cline{3-4} 
                    &                                & min  & max  \\ \hline
1                   & $g_{\sigma}$                              & 6.5  & 15.5   \\
2                   & $g_{\omega}$                              & 6.5  & 15.5   \\
3                   & $g_{\varrho}$                        & 6.5    & 16.5  \\
4                   & $B$                               & 0.5    & 9.0 \\
5                   & $C$                            & -5.0    & 5.0 \\
6                   & $\xi$                                & 0.0 & 0.04  \\ 
7                   & $\Lambda_\omega$                      & 0& 0.12  \\
\hline
\end{tabular}
\end{table}

{\it The Fit Data:-} The fit data, presented in Table \ref{tab1}, include the nuclear saturation density $\rho_0$, the binding energy per nucleon $\epsilon_0$, the incompressibility coefficient $K_0$, and the symmetry energy $J_{\rm sym,0}$, all evaluated at $\rho_0$. Additionally, we incorporate the pressure of pure neutron matter (PNM) at densities of 0.08, 0.12, and 0.16 fm$^{-3}$ from N$^3$LO calculations in chiral effective field theory ($\chi$EFT) \cite{Hebeler2013}, accounting for 2$\times$ N$^3$LO data uncertainty. Furthermore, the likelihood also includes the requirement of the neutron star maximum mass exceeding 2.0 $M_\odot$ with uniform probability.

{\it The Log-Likelihood:-} We optimize a log-likelihood function as a cost function for the given fit data in Table \ref{tab1}. Equation \ref{loglik} represents the log-likelihood function, taking into account the uncertainties $\sigma_j$ associated with each data point $j$ of constraints. The maximum mass of neutron stars is treated differently, using  step function probability,
\begin{equation}
\label{loglik}
    Log (\mathcal{L}) \propto -\sum_j  \left\{ \left(\frac{d_j-m_j(\boldsymbol{\bm{\theta}})}{\sigma_j}\right)^2 + Log(2 \pi \sigma_j^2)\right\} .
\end{equation} 
To populate the seven-dimensional posterior, we employ the nested sampling algorithm \cite{Skilling2004}, specifically the PyMultinest sampler \cite{Buchner:2014nha,buchner2021nested}, which is well-suited for low-dimensional problems. The EoS dataset for subsequent analyses will be generated using the full posterior, which contains 25287 EoS.

\begin{table}[!ht]
\centering
 \caption{The Bayesian inference imposes constraints on various quantities to generate sets of models. These constraints include the binding energy per nucleon $\epsilon_0$, incompressibility $K_0$, and symmetry energy $J_{\rm sym,0}$ at the nuclear saturation density $\rho_0$, each with a 1$\sigma$ uncertainty. Additionally, the pressure of pure neutron matter (PNM) is considered at densities of 0.08, 0.12, and 0.16 fm$^{-3}$, obtained from a $\chi$EFT calculation \cite{Hebeler2013}. The likelihood incorporates a 2$\times$ N$^3$LO uncertainty for the PNM pressure, noting that it increases with density. Furthermore, the maximum mass of neutron stars is constrained to be above 2$M_\odot$.}
  \label{tab1}
 \setlength{\tabcolsep}{5.5pt}
      \renewcommand{\arraystretch}{1.1}
\begin{tabular}{cccc}
\hline 
\hline 
\multicolumn{4}{c}{Constraints}                                                        \\
\multicolumn{2}{c}{Quantity}                     & Value/Band  & Ref     \\ \hline
\multirow{3}{*}{\shortstack{NMP \\  {[}MeV{]} }} 
& $\rho_0$ & $0.153\pm0.005$ & \cite{Typel1999}    \\
& $\epsilon_0$ & $-16.1\pm0.2$ & \cite{Dutra:2014qga}   \\
                               & $K_0$           & $230\pm40$   & \cite{Shlomo2006,Todd-Rutel2005}    \\
                              & $J_{\rm sym, 0}$           & $32.5\pm1.8$  & \cite{Essick:2021ezp}   \\
                              
                               &                 &                &                                                   \\
  \shortstack{PNM \\ {[}MeV fm$^{-3}${]}}                  & $P(\rho)$       & $2\times$ N$^{3}$LO    & \cite{Hebeler2013}   \\
  &$dP/d\rho$&$>0$&\\
\shortstack{NS mass \\ {[}$M_\odot${]}}        & $M_{\rm max}$   & $>2.0$     &  \cite{Fonseca:2021wxt}      \\ 

\hline 
\end{tabular}
\end{table}

\section{Dataset \label{dataset}}
Our goal is to train BNNs models to predict the speed of sound and proton fraction of NS matter from a given set of NS mock observations. To understand the effect of different NS properties on the prediction uncertainty, we generate different datasets using the {25287} EoS that were obtained through the Bayesian analysis formalism (see Sec. \ref{sec:bayesian_setup}). 

\subsection{Structure}
\begin{figure}[!ht]
\includegraphics[width=0.7\linewidth]{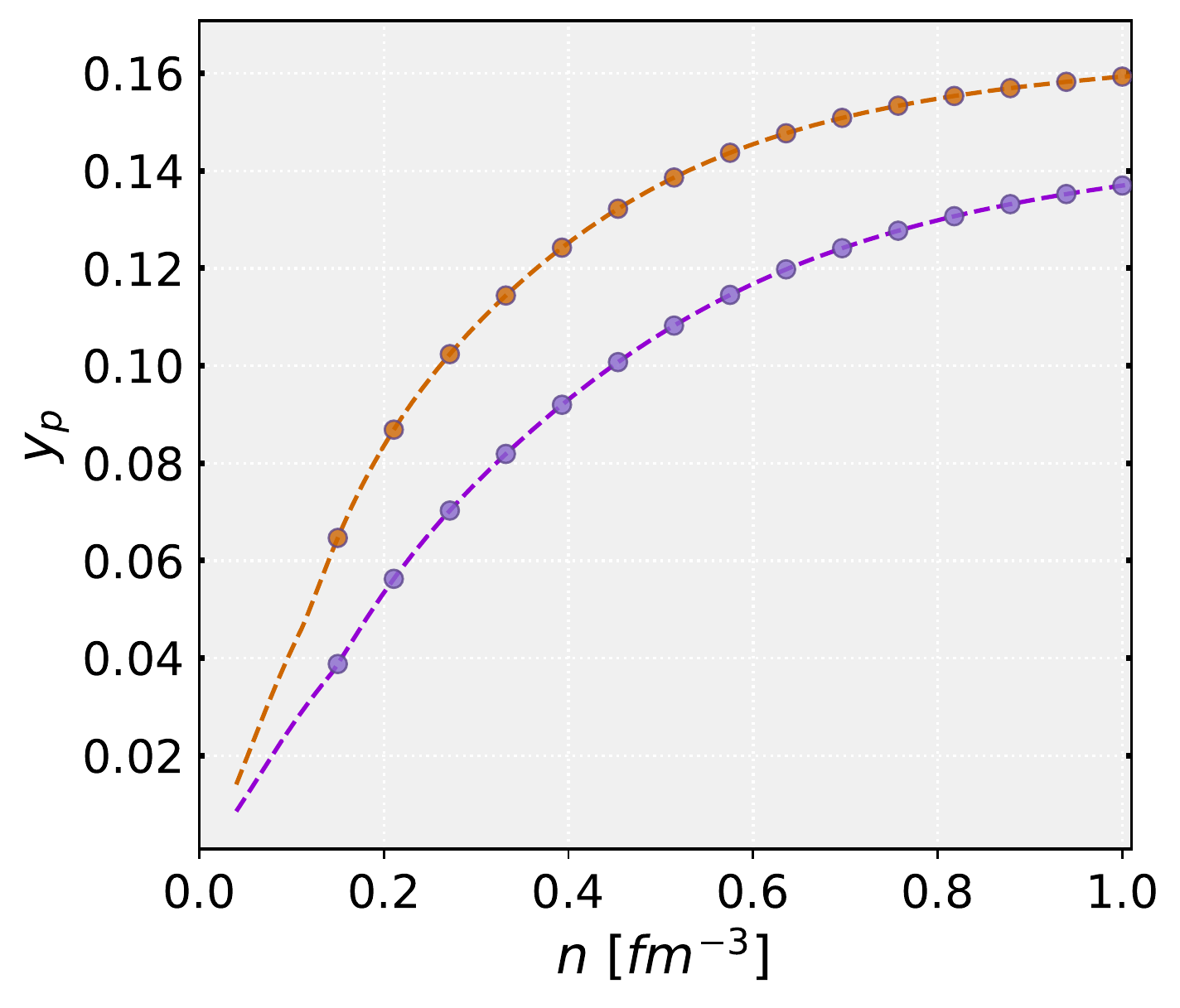}
\caption{Illustration of $\bm{Y}=\bm{y_p}(\bm{n})$ for two EoS: each EoS is represented in our datasets by the 15 points ${y_p}({n_k})$.  }
\label{fig:y_p_points}
\end{figure}

Our BNN model, $P(\bm{Y}|\bm{X},\bm{\theta})$, attributes a probability distribution to $\bm{Y}$ (output space) given a set of NS mock observations $\bm{X}$ (input space), where $\bm{Y}$ denotes the different NS matter properties under study, i.e., the speed of sound $\bm{v_s^2}(\bm{n})$ and proton fraction $\bm{y_p}(\bm{n})$. We have chosen to characterize each element of $\bm{Y}$ at $15$ fixed baryonic densities $n_k$, e.g., $ \bm{y_p}(\bm{n})= [y_p(n_1),y_p(n_2),...,y_p(n_{15})]$. 
The density points are equally spaced between $n_1=0.15$ fm$^{-3}$ and $n_{15}=1.0$ fm$^{-3}$,   $n_k=\{0.15,0.21,0.27,...,1.0 \}$ fm$^{-3}$.
The number of points ($N_Y=15$) was selected as a trade-off between computational training time and the interpolation accuracy, i.e. low residuals between the interpolation and the real values. We illustrate this discretization process of the output space elements for the proton fraction in Fig. \ref{fig:y_p_points}.

Regarding the structure of the input space $\bm{X}$, two different element structures are studied: i) $ \bm{X}= [M_1,...,M_5,R_1,...,R_5]$ corresponding to five $M_i(R_i)$ simulated observations and ii) $ \bm{X}= [M_1,...,M_5,R_1,...,R_5,M_1^{'},...,M_5^{'},\Lambda_1,....,\Lambda_5]$ corresponding to five $M_i(R_i)$ and five $\Lambda_j(M_j^{'})$ 
 simulated observations. In summary, the output elements $\bm{Y}_i$ of our datasets are specified by 15-dimensional vectors and 
the input space elements $\bm{X}_i$ by 10-dimensional or 20-dimensional vectors, depending on the dataset type under study. The statistical procedure for generating the different synthetic observational datasets is presented in the following. 

\subsection{Generation\label{generation}}
The first step of the generation of the datasets consists in randomly splitting the total number of EoS into train and test sets
in a proportion of {80\%/20\%}, i.e., the train set contains {22758 EoS} while the test has {2529 EoS}. Secondly, we generate two types of datasets that share the $\bm{Y}_i$ structure but with different $\bm{X}_i$ structures:
\begin{align*}
    \bm{X_i}&= [M_1,...,M_5,R_1,...,R_5]\\
    \bm{X_i^{'}}&= [M_1,...,M_5,R_1,...,R_5,M_1^{'},...,M_5^{'},\Lambda_1,....,\Lambda_5].
\end{align*}
These different structures allows us to compare and access
how informative is the tidal deformability on the  model predictions.
The statistical generating procedure is composed of the following steps. For each EoS, we randomly select 5 NS mass values, $M_i^{(0)}$, from a uniform distribution between $1.0M_{\odot}$ and $M_{\text{max}}$. Then, the radius $R_i$ is sampled from a Gaussian distribution centred at the TOV solution, denoted as $R(M_i^{(0)})$, and with a standard deviation of $\sigma_R$. Finally, we sample the final NS mass from a Gaussian distribution centered at $M_i^{(0)}$ and a standard deviation of  $\sigma_M$. The above process can be summarized in the following equations:
 \begin{align}
   M_i^{(0)} &\sim \mathcal{U}{[1,M_{\text{max}}]}\quad (\text{in units of  }\textup{M}_\odot)\\
        R_i &\sim \mathcal{N}\left(R\left(M_i^{(0)}\right),\sigma_R^2 \right) \\
    M_i &\sim \mathcal{N}\left(M_i^{(0)},\sigma_M^2 \right),  \quad i=1,...,5  
 \end{align}
 
The final generated elements consists of $\bm{X_i}=[M_1,...,M_5,R_1,...,R_5]$, and each one is a possible realization ({\it observation}) that characterizes the $M(R)$ diagram of the specific EoS. 
This procedure is similar to the one used in \cite{fujimoto2021extensive}, where a Gaussian noise was applied to 15 values from $M(R)$ curve, and then shifted from the original mass radius curve: $M_i = M_i^{(0)} + \mathcal{N}\left(0,\sigma_M^2 \right)$ and $R_i ={R}\left(M_i^{(0)}\right)+ \mathcal{N}\left(0,\sigma_R^2 \right)$, for $i=1,...,15$.   The second kind of datasets includes the tidal deformability and has the additional steps
 \begin{align}
    M_j^{'} &\sim \mathcal{U}{[1,M_{\text{max}}]}\quad (\text{in units of  }\textup{M}_\odot)\\
    \Lambda_j &\sim  \mathcal{N} \left( \textbf{$\Lambda$} ( M_j^{'}  ),\sigma_{\Lambda}^2( M_j^{'}  ) \right)  \quad j=1,...,5,
 \end{align}
where $\Lambda ( M_j^{'} )$ is given by the $\Lambda(M)$ relation of the specific EoS and $\sigma_{\Lambda}( M_j^{'} )$ describes an overall dispersion around the mean value. 
Samples values with $\Lambda<0$ were discarded.
The functional form $\sigma_{\Lambda}( M_j^{'} )$ should reflect our expectation on the mock observational uncertainty as a function of the NS mass. In the present work, we considered $\sigma_{\Lambda}( M_j^{'} )=\text{constant}\times \hat{\sigma}(M_j^{'})$, where $\hat{\sigma}(M_j^{'})$ is the standard deviation of $\Lambda(M)$ determined from dataset of EoSs.
The generated point is $\bm{X_i}=[M_1,...,M_5,R_1,...,R_5,M_1^{'},...,M_5^{'},\Lambda_1,....,\Lambda_5]$ (a similar approach can be found in \cite{Traversi_2020}). 
In the above procedures, there is an additional parameter, which we denote by $n_{\text{s}}$, that specifies the number of mock observations for each EoS, i.e., the number of times the above procedures are applied to each EoS. For instance, choosing $n_s=20$  would mean running the above procedures 20 times for each EoS (20 {\it observations}), and thus obtaining   $\chav{\bm{X_1},\bm{X_2},...,\bm{X_{20}}}$.\\

\begin{figure}[!ht]
\includegraphics[width=1\linewidth]{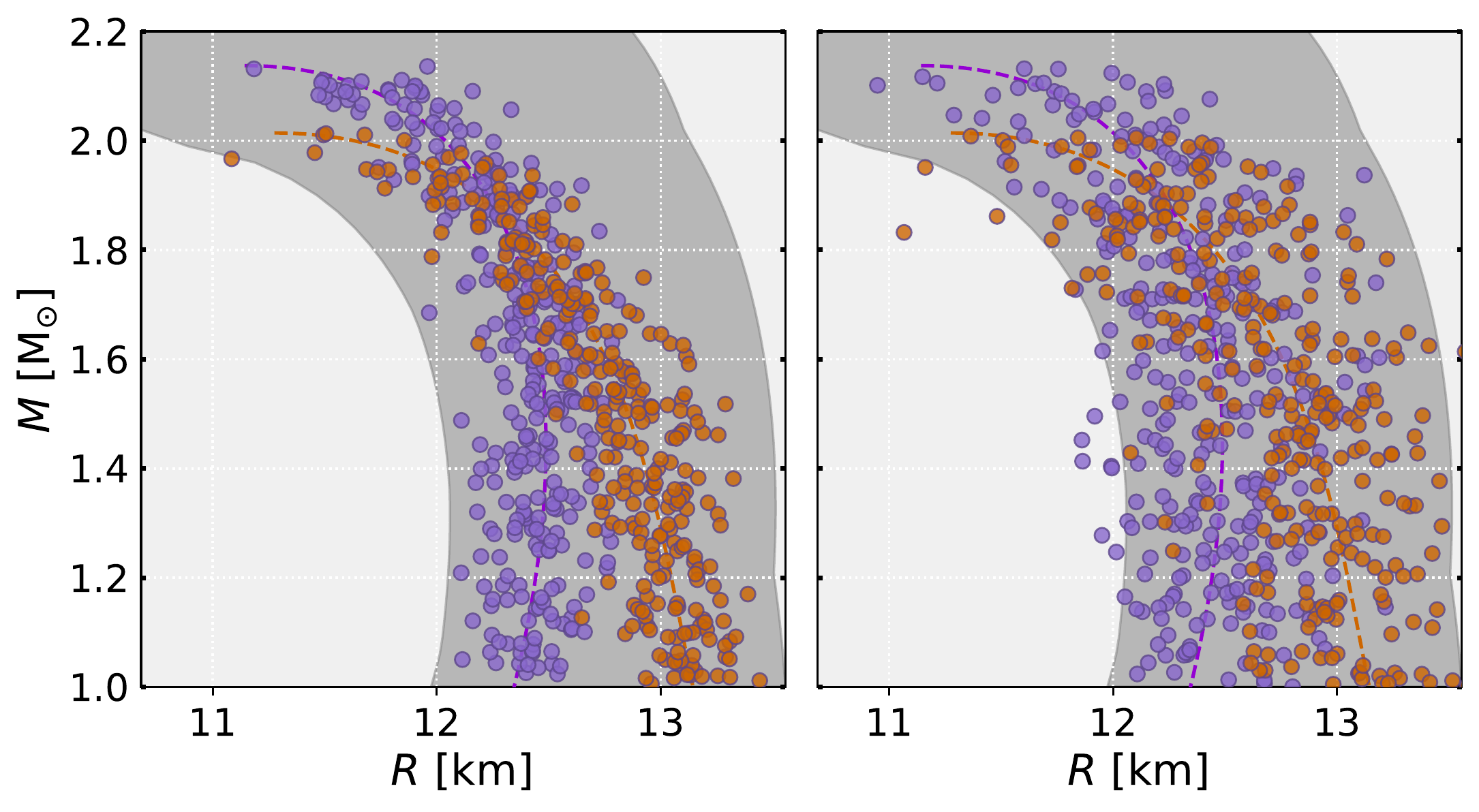}
\caption{The $n_s=60$ mock observations generated for two EoSs in dataset 1 (left) and dataset 2 (right). The grey area represents the extremes of our EoS dataset. The two EoS coincide with the ones used in Fig. \ref{fig:y_p_points}.}
\label{fig:MR_set1_2}
\end{figure}

\begin{figure}[!ht]		
\includegraphics[width=1\linewidth]{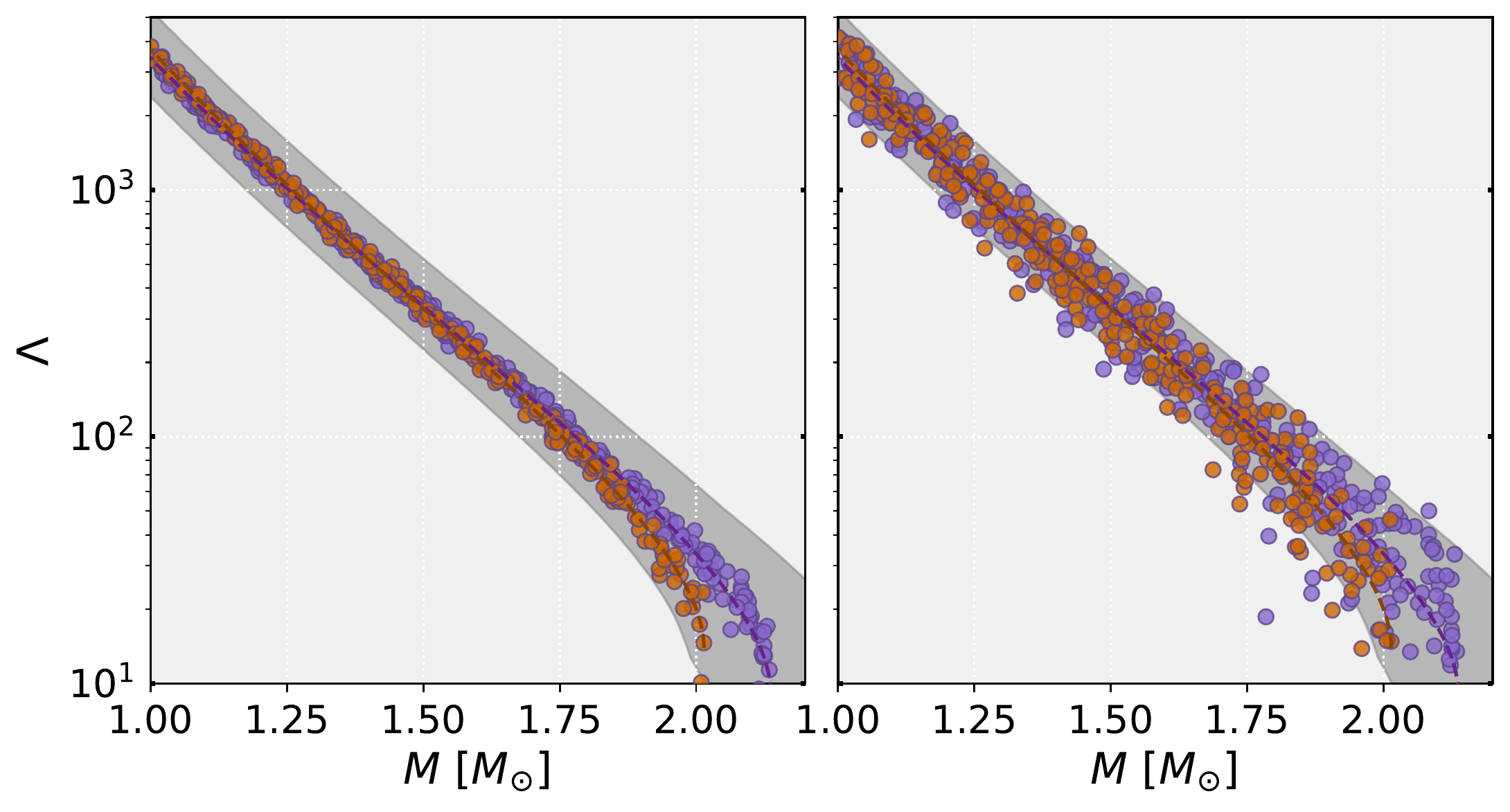} 
\caption{The $n_s=60$ mock observations generated in the $\Lambda-M$ diagram for two EoSs in dataset 3 (left) and dataset 4 (right). The grey area represents the extremes of our EoS dataset. The two EoS coincide with the ones used in Fig. \ref{fig:y_p_points}.}
\label{fig:Lambda_set3_4}
\end{figure}

Applying the above formalism to both the train and test sets, we have generated a total of 4 datasets whose properties are displayed in Table \ref{tab:sets}. Sets 1 and 2 only contain information about the NS radii (input space $\bm{X}$ is 10-dimensional) while sets 3 and 4 also include the tidal deformability (input space $\bm{X}$ is 20-dimensional). The analysis of sets 1 and 2 allows to understand how a decrease in the spread of the mock observations around the TOV solution affects the predictions and uncertainties. In the same manner, sets 3 and 4 aim to understand possible effects on the model predictions arising from an increase of simulated observations scattering around the mean value on the tidal deformability. We use 60 mock observations, $n_s=60$, on the training sets for each EoS while $n_s=1$ was employed for the test sets. This key difference tries to simulate a real case scenario in which we only have access to a single mock observation of the {\it true} EoS. Here, by single mock observation, we mean $n_s=1$ that corresponds to five $M_i(R_i)$ mock observations (sets 1 and 2) or five $M_i(R_i)$ mock observations and five $\Lambda_j(M_j)$ mock observations (sets 3 and 4). To illustrate the dataset generation, Fig. \ref{fig:MR_set1_2} displays the 60 mock observations for two distinct EoSs that belong to the generated datasets (dataset 1 and 2 on the right and left figures, respectively). Note that, for each EoS, there are 300 points in the $M(R)$ diagram, i.e.,  consisting of 60 EoS simulated observations for 5 NS mock observations each.  It is clear the increase of $\sigma_R$ from dataset 1 to 2, highlighting the differences between the two datasets. Furthermore, we provide Figure \ref{fig:Lambda_set3_4} to depict the tidal deformability values for datasets 3 and 4.

\begin{table}[!ht]
   \caption{Generation parameters for each dataset. $\hat{\sigma}(M_j)$ denotes the standard deviation of $\Lambda(M)$ calculated on the train set. }
    \label{tab:sets}
    \begin{tabular}{cccc}
    \toprule
    Dataset & $\sigma_M \;[ M_\odot]$ & $\sigma_R \;$ [km]   & $\sigma_{\Lambda}( M_j )$ \\ \hline
    1  & 0.05       & 0.15       &       ---        \\ \hline
    2  & 0.1        & 0.3       &        ---   \\ \hline
    3  &  0.1   & 0.3    & 0.5$\hat{\sigma}(M_j)$               \\ \hline
    4  &  0.1   & 0.3 & 2$\hat{\sigma}(M_j)$               \\ \hline
    \end{tabular}
\end{table}

\subsection{Training procedure}
\label{subsection}

To assess the response of Bayesian Neural Networks (BNNs) to varying input noises and output targets, we have conducted experiments involving the training of diverse functional and stochastic models (as explained in Section \ref{bnn}).
These BNN models were trained using distinct datasets generated as outlined in Section \ref{generation}.  
During the training stage, a subset of the training data was randomly selected as validation step, i.e., the training data was split into 
80\% actually for training and 20\% for validation. Moreover, the input data $\bm{X}$ was standardized.\\

Defining the functional models involves adjusting the number of neurons, layers, and activation functions. Table \ref{tab:final_models2} shows the best functional models for each dataset mentioned in Table \ref{tab:sets}. For the hidden layers, we explore hyperbolic tangent, softplus, and sigmoid activation functions, while utilizing a linear activation function for the output layer.
As the input vector sizes differ (10  for sets 1 and 2, and 20 for sets 3 and 4), we employ more neurons per layer for larger input spaces. This is due to the increased complexity demanded by a greater number of parameters. Specifically, we use 15 and 10 neurons for sets 1 and 2, and 20 and 25 neurons for sets 3 and 4, respectively. 
The output layer consistently contains 30 neurons, with 15 representing the mean and 15 representing the standard deviation of the output probability distribution function. It is worth noting that we deliberately excluded the use of correlation in the output layer due to the inferior performance observed when attempting to incorporate it. As a result, the output layer is solely focused on capturing the mean and standard deviation information of the output distribution.  The architecture employed in this study involved utilizing two to three hidden layers. The number of neurons within each hidden layer remained consistent, but it varied depending on the size of the input vector, as explained earlier. During the hyper-parameter search process, we systematically explored four different architectures for each output. However, we narrowed our focus to datasets 1 and 3, as we specifically aimed to identify the most suitable architecture for the two different input sizes.
The best outcomes are obtained by employing sigmoid as the activation function in the hidden layers, ensuring minimal loss and preventing divergence. 
Across all eight dataset configurations, i.e., two outputs ($v_s^2$ and $y_p$) and four datasets, see Table \ref{tab:sets}, it was found that the optimal number of hidden layers is two.  For sets 1 and 2, the best performance was achieved with 15 neurons in each hidden layer, while for sets 3 and 4, 25 neurons were utilized in each hidden layer. Detailed information on these configurations can be found in Table \ref{tab:final_models2} for the two output variables. 
During training, we employ a learning rate of 0.001 and utilize the ADAM optimizer \cite{kingma2014adam} with the AMSgrad improvement \cite{reddi2019convergence}. The models are trained for 4000 epochs, with a mini-batch size of 768. \\

\begin{table}[ht!]
\caption{Structures of the final BNN models.
The $v_s^2(n)$ and $y_p(n)$ models have the same structure.}
\centering
\begin{tabular}{c|c|cc}
\multirow{2}{*}{\textbf{Layers}} & \multirow{2}{*}{\textbf{Activation}} & \multicolumn{2}{c}{\textbf{Neurons}} \\ \cline{3-4} 
 &  & \multicolumn{1}{c|}{\textbf{Dataset 1 \& 2}} & \textbf{Dataset 3 \& 4} \\ \hline
Input & N/A & \multicolumn{1}{c|}{10} & 20 \\ \hline
Hidden Layer 1 & Sigmoid & \multicolumn{1}{c|}{15} & 25 \\ \hline
Hidden Layer 2 & Sigmoid & \multicolumn{1}{c|}{15} & 25 \\ \hline
Output & Linear & \multicolumn{1}{c|}{30} & 30 \\ \hline
\end{tabular}
\label{tab:final_models2}
\end{table}

Regarding the stochastic model, we adopt a Gaussian prior with mean zero and standard deviation of one as mentioned in Section \ref{sec:vi}. While this prior choice lacks a specific theoretical justification, it serves as a reasonable default prior, as discussed in \cite{jospin2022hands}. Future research could delve further into investigating the impact of prior parameters, similar to the approach taken in reference \cite{bollweg2020deep}. Additionally, we select a multivariate normal distribution as the variational posterior as explained in Section \ref{sec:vi}, initialized with mean 0 and a diagonal covariance matrix, where the standard deviation is equal to $\log(1+\exp{0}) =0.693$. 
Furthermore, we opt for a deterministic output layer instead of a probabilistic one, as it has demonstrated improved results in our experiments. This decision is motivated by the fact that the deterministic output layer aligns better with the specific requirements of our model architecture and the nature of the problem we are addressing. All BNNs models were coded using Tensorflow library \cite{tensorflow2015-whitepaper}, more specifically we use Keras \cite{chollet2015keras}, an high-level API of the TensorFlow.

\section{Results: neutron star matter properties \label{results}}

In the following, we discuss the results for the speed of sound squared $v_s^2(n)$ and proton fraction $y_p(n)$. To analyze how the observational uncertainty in $R$ and $\Lambda$ affects the model confidence, we are going to compare the results from the different models of Tab. \ref{tab:final_models2}, which were trained on the different datasets presented in Tab. \ref{tab:sets}.
For the sake of simplicity, whenever we want  to distinguish the different 8 BNNs models, we will, hereafter, refer to their training (data)sets, 1 to 4, and predicting quantity, $v_s^2(n)$ or $y_p(n)$.
For instance, when we say the results of dataset 2 for $y_p(n)$, we mean the BNN model trained on dataset 2 that has the proton fraction as the target quantity.

\subsection{Speed of sound \label{results_vs}}

\begin{figure}[!ht]		
\includegraphics[width=.95\linewidth]{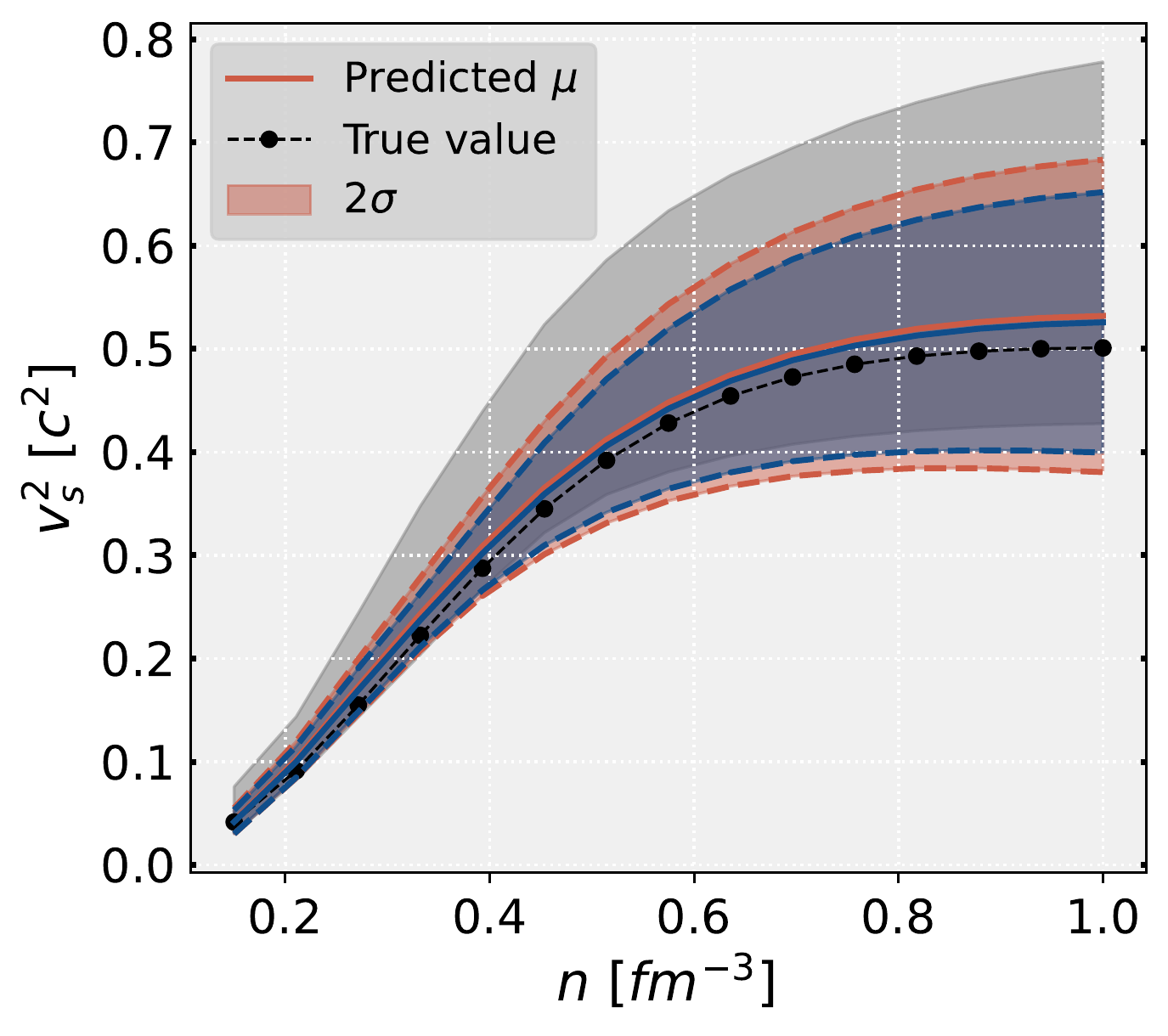} \\
\includegraphics[width=.95\linewidth]{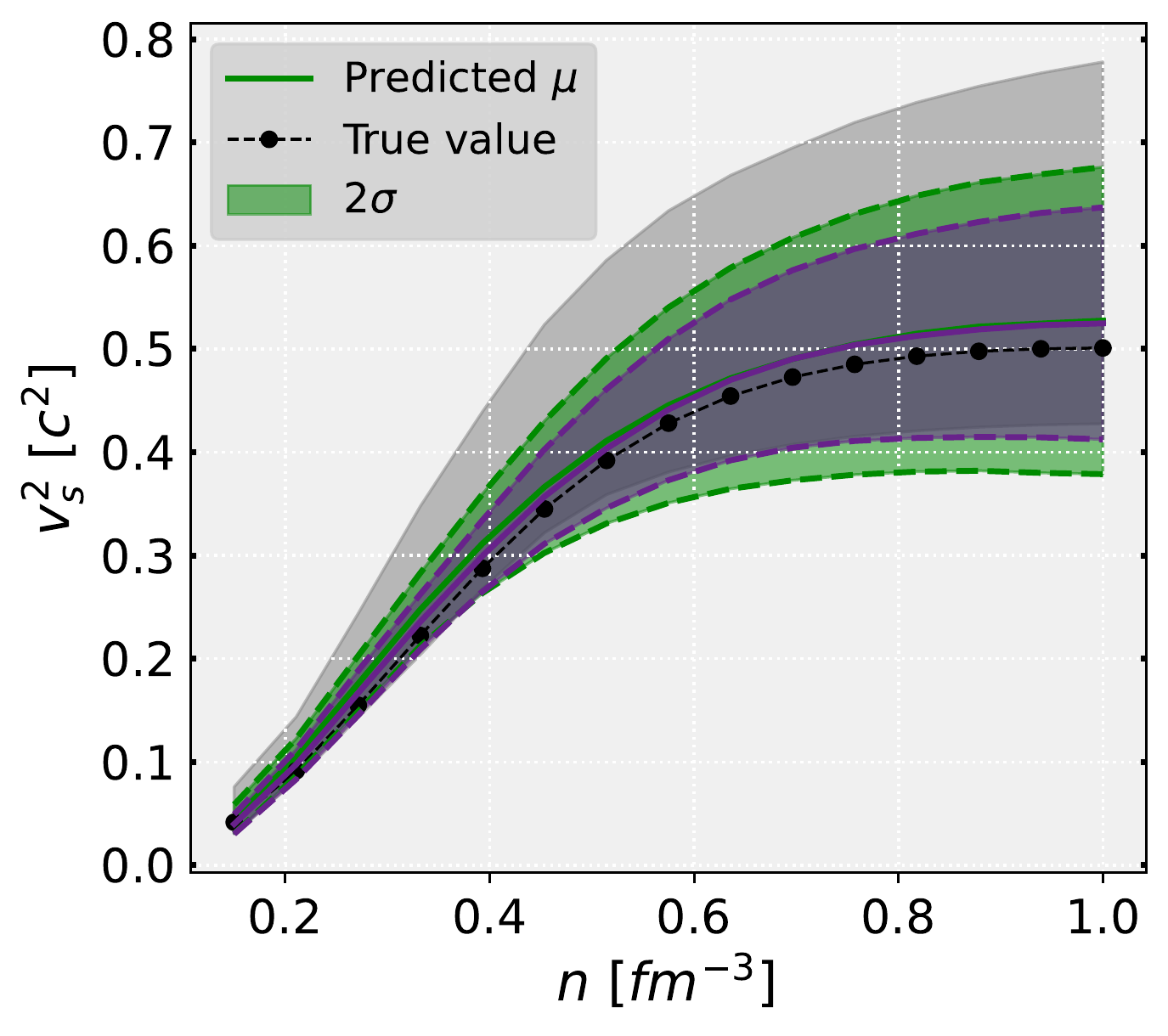} 
\caption{The BNNs predictions for $v_s^2(n)$ using one EoS of the test. The models trained on datasets 1 (blue) and 2 (orange) are in the upper figure while datasets 3 (purple) and 4 (green) models are in the lower figure. The prediction mean values (solid lines) and $2\sigma$ confidence intervals are shown. The true values are shown in black dots and the range of $v_s^2(n)$ from the train set is indicated by the grey region. }
\label{fig:vs_distr}
\end{figure}

Let us start by illustrating the different BNNs predictions for the speed of sound squared on a randomly selected EoS from the test set. The results are displayed in Figure \ref{fig:vs_distr}, where the top panel shows sets 1 (blue) and 2 (orange) models while the below figure shows sets 3 (purple) and 4 (green). Model predictions are computed using Eq. \ref{eq:baye_pred}, from which we show the mean values (solid lines) and $2\sigma$ regions (color regions). The upper plot clearly shows that the prediction uncertainty, characterized by the distribution's standard deviation $\sigma$, is smaller on set 1 than on set 2, and most importantly, the predicted mean values are close to the real values. The same pattern is seen in the lower figure: set 3 model (purple) has a lower prediction variance than set 4. While a deeper understanding of the overall behaviour requires that we analyze the whole test set, Fig. \ref{fig:vs_distr} already indicates that the BNNs models are able to capture the characteristics of the different datasets (see Table \ref{tab:sets}): the increased dispersion of the NS mock observations around the true values translate into a larger uncertainty when inferring the corresponding EoS properties. \\

\begin{figure}[!ht]		
\includegraphics[width=.95\linewidth]{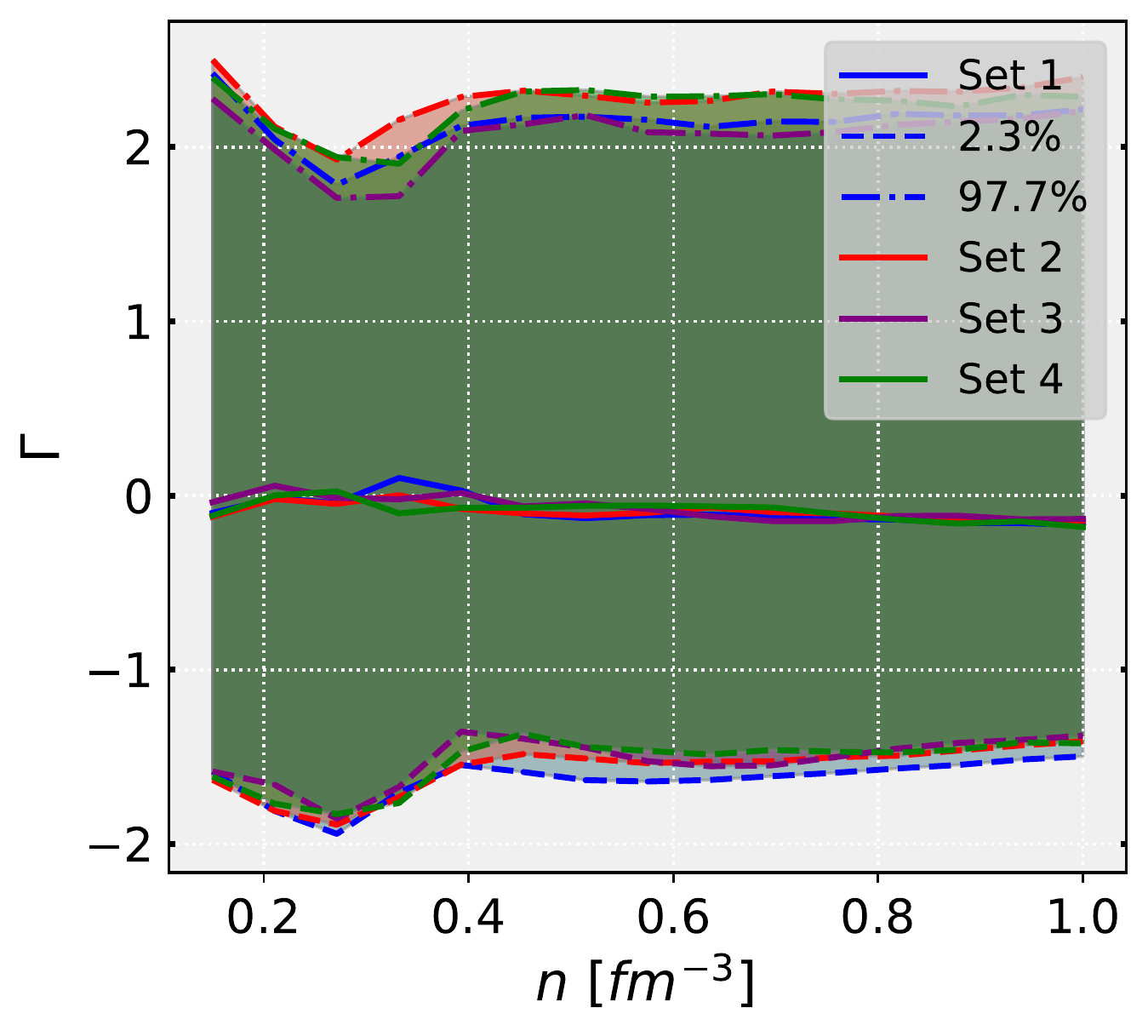} \\
\includegraphics[width=.95\linewidth]{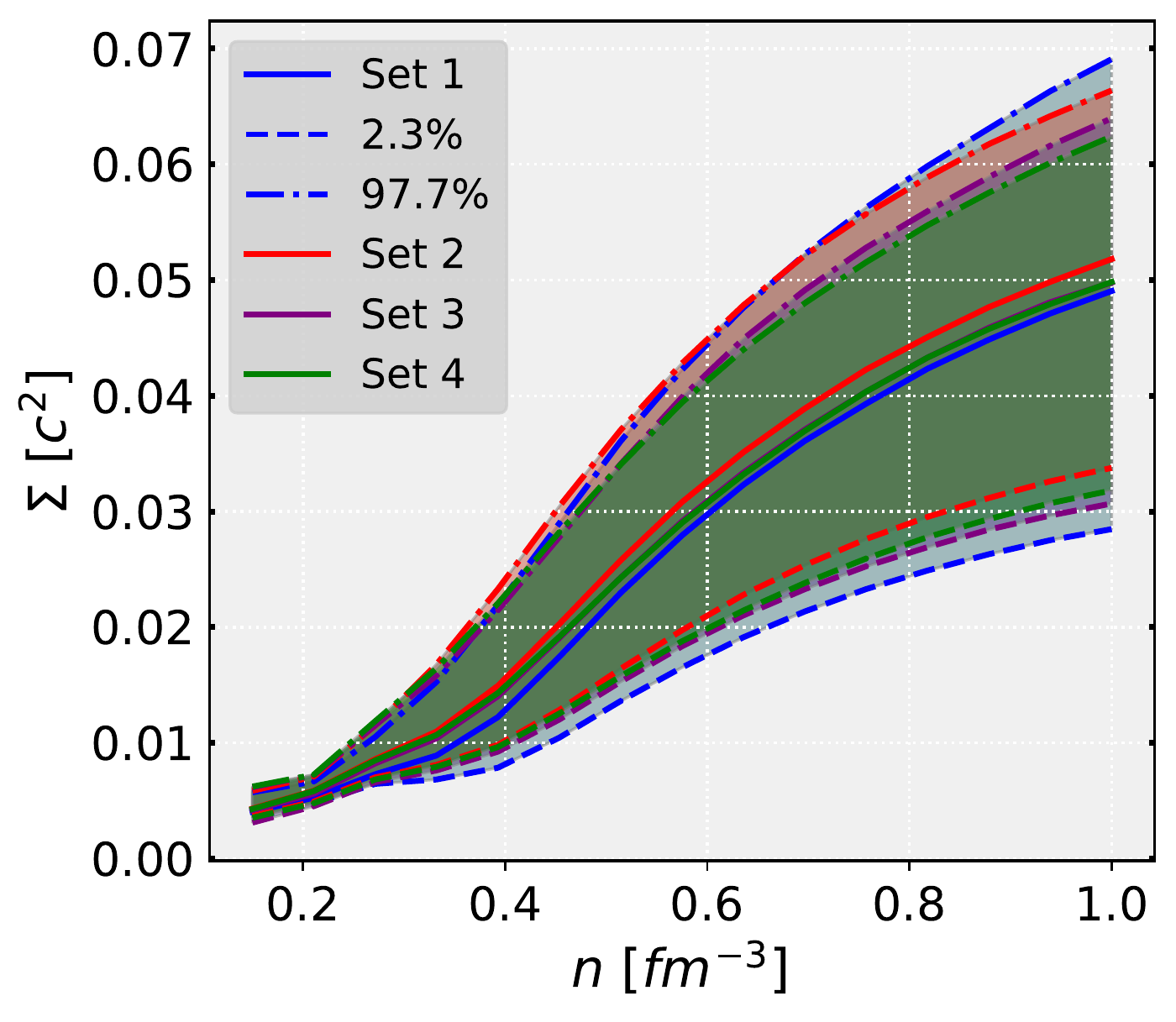} 
\caption{Median (solid line), 95.4\% confidence interval (dashed and dotted lines), and extreme values (region) for $\Gamma(n_k)=\pc{v_s^2(n_k)-v_s^2(n_k)^{\text{true}}}/\sigma(n_k)$ (top) and $\Sigma(n_k)=\sigma(n_k)$ (bottom) for each dataset. }
\label{fig:vs_sd}
\end{figure}

To investigate how the models predictions behave over the entire test set, we define the normalized residuals' predictions as $\Gamma(n_k)=\pc{v_s^2(n_k)-v_s^2(n_k)^{\text{true}}}/\sigma(n_k)$ and the dispersion by $\Sigma(n_k)=\sigma(n_k)$ at each of the prediction densities, i.e., $k=1,...,15$. A summary statistics of both quantities over all EoS of the test set is shown in Fig. \ref{fig:vs_sd}.
The distribution of $\Gamma(n)$ (top panel), for all 4 datasets, has 50 \% of the values near zero, indicating that the median of the prediction values is unbiased. Furthermore, at the 2.3 \% and 97.7\% of the cumulative percentage, we can see that the distribution lies between -2$\sigma$ and 2$\sigma$ respectively, indicating that the prediction mean deviates from the true value less than $2\sigma$ 95.4\% of the times.
The fact that the distribution properties of $\Gamma(n)$ are similar across all datasets and independent of the density reveals that the BNNs models are correctly modelling the dispersion of the predictions considering the corresponding mean residuals at each density value. The value of $\Sigma(n)$ (bottom panel) grows with increasing density reflecting the training set statistics: the EoS dataset was generated by Bayesian inference where saturation properties were imposed, leading to a wider uncertainty at higher densities while low density regions are strongly constrained.
From the $\Sigma(n)$ plot, we see that the whole distribution of the BNN model trained on set 1 (blue line) shifts to lower values, specifically on $2.3\%$ and $50\%$ of the values, showing that there is a considerable decrease in uncertainty when the dispersion of NS mock observations is reduced by a factor of two, from $(\sigma_M = 0.1 M_{\odot},\sigma_R=0.3\text{ km})$ to $(\sigma_M = 0.05 M_{\odot},\sigma_R=0.15\text{ km})$ (see Table \ref{tab:sets}).\\

\begin{figure}[!ht]		
\includegraphics[width=.75\linewidth]{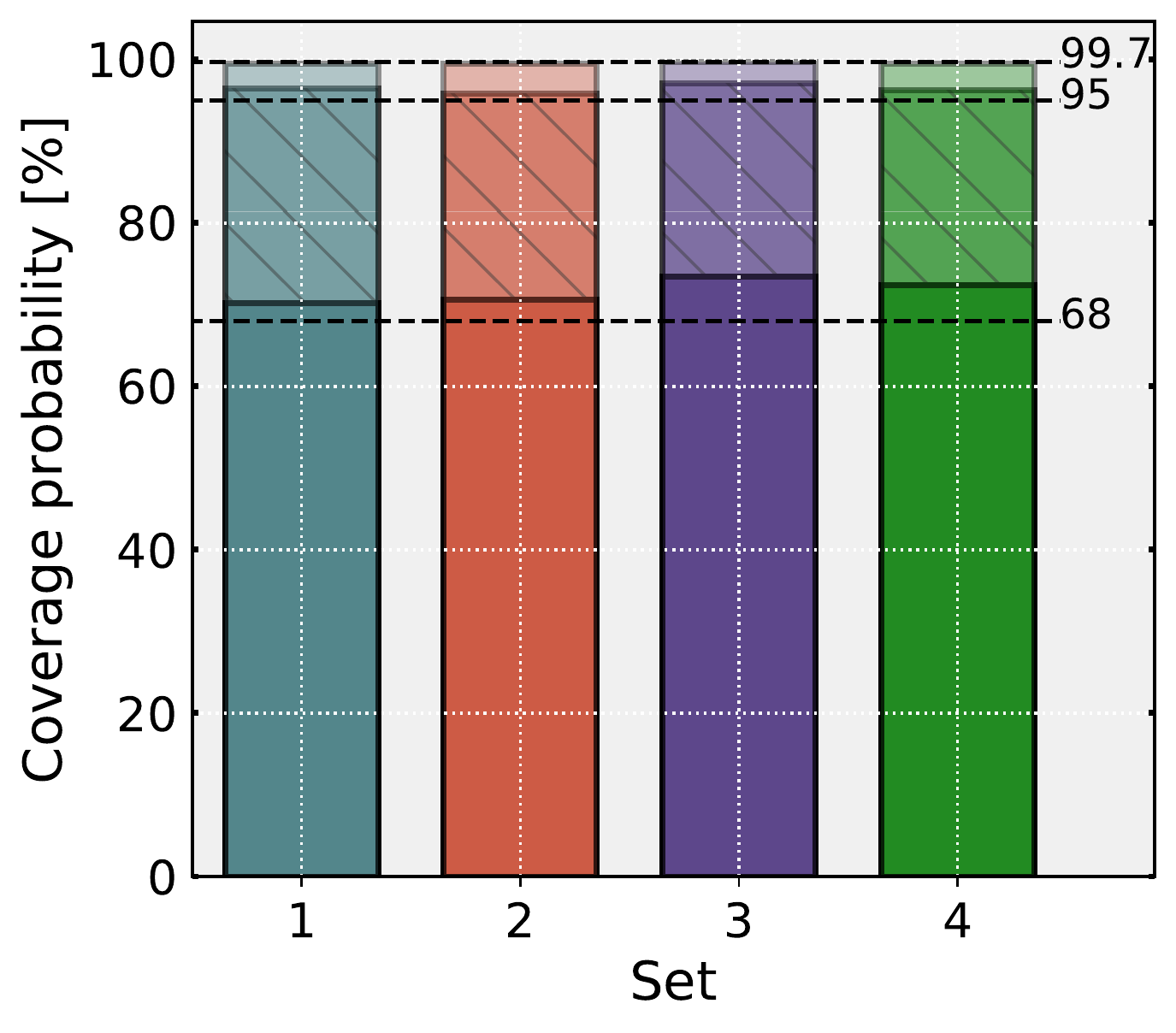} 
\caption{Coverage probability calculated on the test set of the $v_s^2(n)$ BNNs models.}
\label{fig:vs_cover}
\end{figure}

To estimate the overall performance of the different BNNs models, we show the coverage probability of each model  in Fig. \ref{fig:vs_cover}. 
The coverage probability quantifies how the model is perceiving the distribution of the data, by determining if the percentage of values contained in $1\sigma$ of the output distribution, i.e., the number of values in a specific interval divided by the total number of values, corresponds to {68.3\%} of the number of values we  are using for the test set. The same analogy is repeated for $2\sigma$ (95\%) and $3\sigma$ (99.7\%), this was implemented for each of the 15 values of the output independently, obtaining the respective coverage probabilities relative to the output densities, and then we estimated the mean of this 15 coverage probabilities
represented in Fig. \ref{fig:vs_cover}  for the 4 sets. The overall results  show that the model is correctly estimating the distribution of the data, since the bars are very close to the three percentages, a little fluctuation can be seen in  the {68.3\%}, where the 4 sets are overestimating the uncertainty of the results, meaning, we have a percentage of data bigger than {68.3\%}  in 1$\sigma$ of the model, so the $\sigma$ of the model should be smaller, this can be seen even more for set 3. \\

\begin{figure}[!ht]		
\includegraphics[width=.95\linewidth]{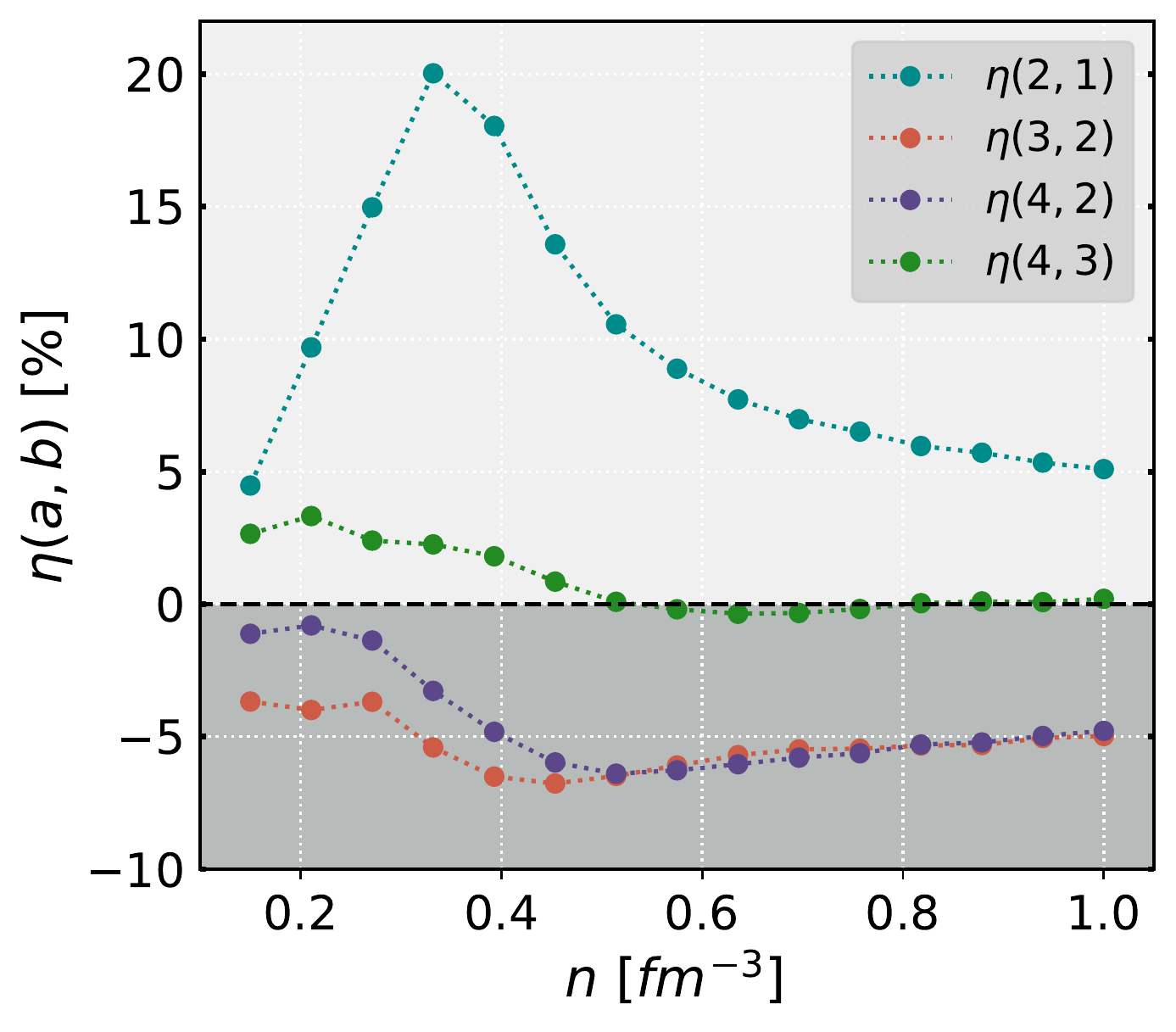}
\caption{Prediction uncertainty deviation $\eta\pr{a,b}$ between the $v_s^2(n)$ BNN models $a$ and $b$ (see text for details). }
\label{fig:vs_eta}
\end{figure}

To quantify how the increase in the mock observational scattering of $(R,\Lambda)$ affects the model prediction uncertainties, let us define the following quantity 
\begin{equation}
  \eta\pr{a,b}(n_k)=\frac{1}{T}\pc{\sum_{i=1}^{T}\frac{\sigma_i^{b}(n_k) - \sigma_i^a(n_k) }{\sigma_i^a(n_k)}} \times 100,  
  \label{eq:eta}
\end{equation}
where $T$ is the total number of EoS in the test set, and $k=1,...,15$. This quantity defines the percentage uncertainty deviation between models $b$ and $a$ at density $n_k$. Figure \ref{fig:vs_eta} shows the results, where we plotted 4 different comparisons. The first conclusion, looking at the result for $\eta\pr{2,1}$ (cyan), is that the prediction uncertainty increased when we consider the BNN model of dataset 1 compared with the one from dataset 2. Furthermore, the $\eta\pr{2,1}$ reached the maximum value of 20\% at $n_4=0.33$ fm$^{-3}$. In other words, when the (synthetic) observational data scattering doubles, from $(\sigma_M = 0.05 M_{\odot},\sigma_R=0.15\text{ km})$ to $(\sigma_M = 0.1 M_{\odot},\sigma_R=0.3\text{ km})$, the uncertainty increases of the order 5-20\%, having the largest value at the densities where the NS radius, $R(M)$, and $v_s^2(n)$ have the highest correlation (see Annex \ref{append_1} for details and ref. \cite{Ferreira:2019bgy}).  The region of larger sensitivity to the uncertainty of the mock observational data coincides with the density interval where the speed of sound increases steadily, and, in many agnostic approaches attains a maximum  followed by a decrease or flattening  at larger densities \cite{Altiparmak:2022bke,Gorda:2022jvk,Somasundaram:2021clp}. The nucleonic EoS that has been used to train the model also shows similar behavior for the speed of sound \cite{Malik:2023mnx}. This behavior of the speed of sound for densities below three times saturation density is dictated by the two solar mass constraints. 

The second important conclusion is the impact of adding the $\Lambda$ information on the inference properties. This point is clear when analysing the dependence and values of $\eta\pr{3,2}$ (the difference between datasets 2 and 3 is that the latter contains information on the tidal deformability, see Tab. \ref{tab:sets}). The negative values reflect the fact that the prediction uncertainty decreases when the tidal information is added to the training procedure -- the tidal deformability is informative of the  $v_s^2(n)$ of neutron star matter. The maximum value of the uncertainty decrease is 7\%, occurring at $n_6=0.45$ fm$^{-3}$. 
Similarly, when considering $\eta\pr{4,2}$ (blue), it also exhibits negative values throughout, indicating a decrease in uncertainty compared to dataset 2. However, beyond the seventh density, the two values become very close to each other, suggesting that no additional information is being perceived by including more dispersion on the tidal deformability.
Looking at  $\eta\pr{4,3}$ and comparing it with  $\eta\pr{2,1}$, we see that for the first density, they differ very little, but from there on $\eta\pr{2,1}$ is approximately always more than 5 times bigger.  Noting an important consideration here, the proportion of input values being altered: dataset 3 and 4 involves changing only the uncertainty of 5 quantities out of the 20  input values, i.e. changing  a quarter of our input vector, while in dataset 1 and 2 we have changed all the input values. So we could anticipate an at least four-fold increase in $\eta$ for datasets 1 and 2, however, this percentage is the majority of times even bigger. This implies that the dispersion of the mass-radius pairs has a more significant impact on the model compared to the mass-tidal deformability pairs. By acknowledging this difference in the proportion of modified input values, we gain insight into how the model's understanding and response to dispersion changes are influenced. \\

\subsection{Proton fraction\label{results_yp}}

\begin{figure}[!ht]		
\includegraphics[width=.95\linewidth]{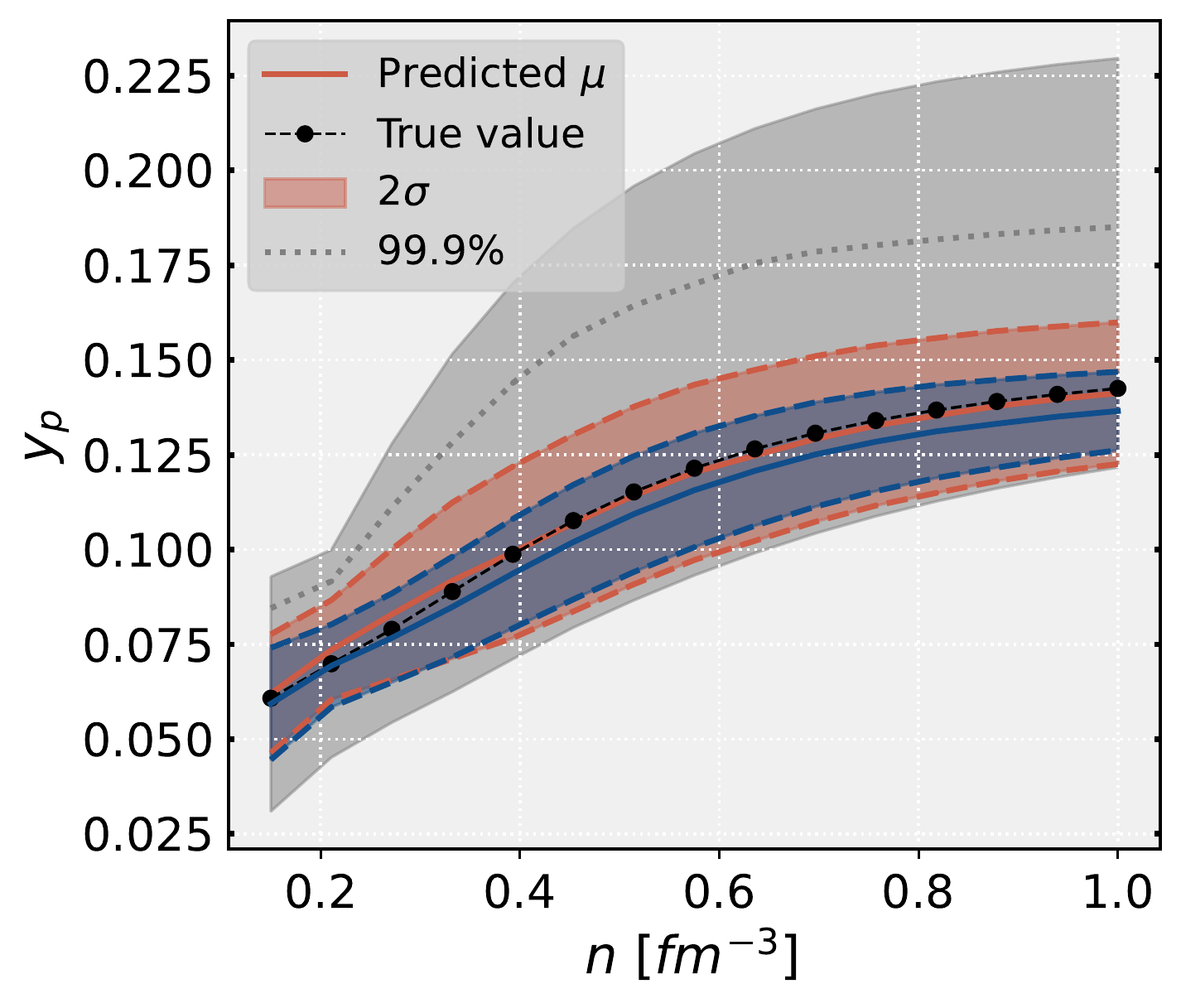} \\
\includegraphics[width=.95\linewidth]{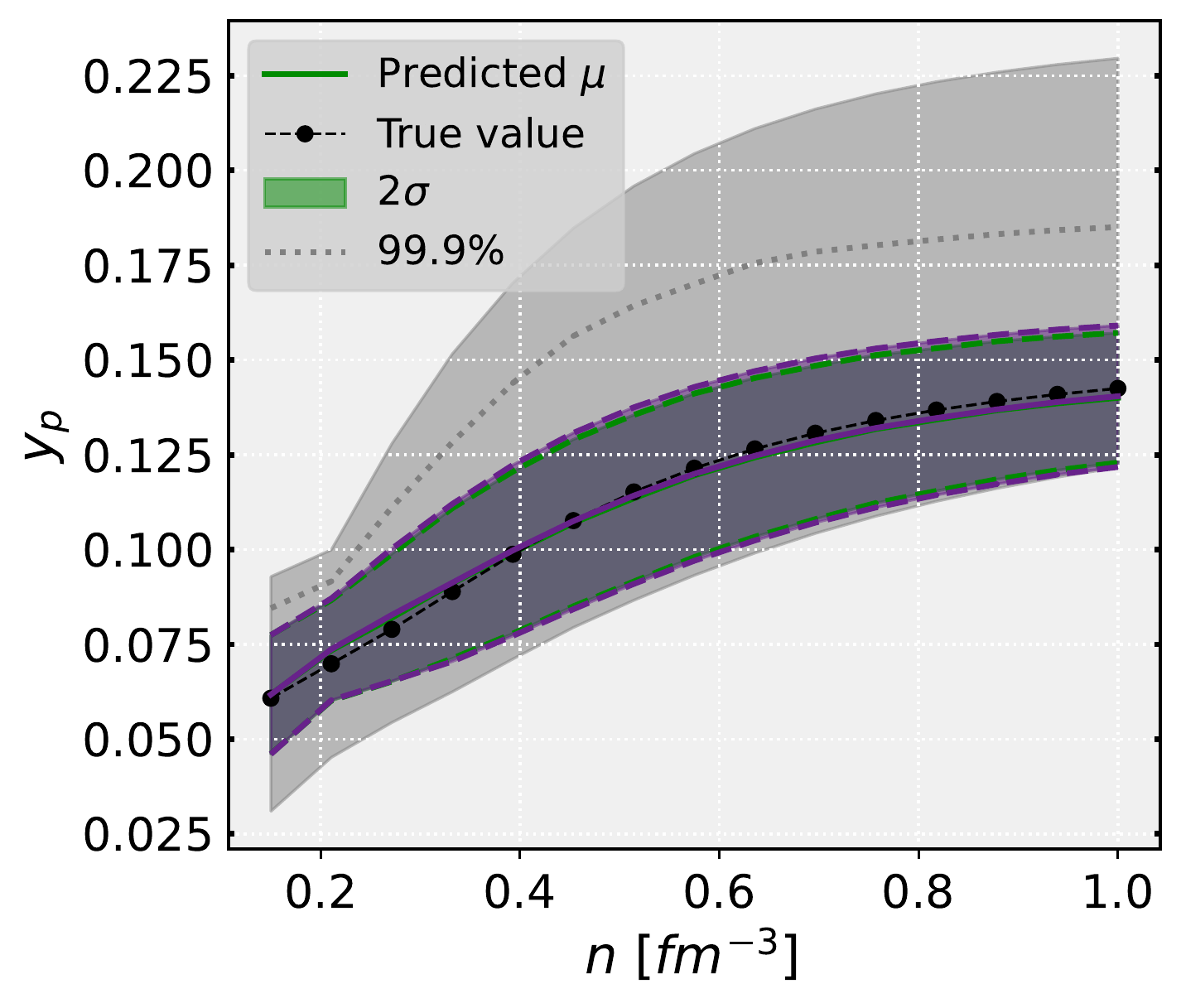} 
\caption{The BNNs predictions for $y_p(n)$ using one EoS of the test. The models trained on datasets 1 (blue) and 2 (orange) are in the upper figure while datasets 3 (purple) and 4 (green) models are in the lower figure. The predicted mean values (solid lines) and $2\sigma$ confidence intervals are shown. The true values are shown in black dots.}
\label{fig:yp_distr}
\end{figure}

Let us now analyze the model's predictions for the proton fraction $y_p(n)$. Using a specific EoS from the test dataset (for illustration purposes), we show in Fig. \ref{fig:yp_distr} the models' prediction for each dataset: 1 (blue) and 2 (orange) on the upper panel and 3 (purple) and 4 (green) in the lower panel. The range of $y_p(n)$ from the train set is indicated by the grey region and the dashed grey line displays the $99.9\%$ percentage of data/probability line. The train statistics is important to point out that the upper region between the  $99.9\%$ probability and maximum boundary lines is caused by the presence of just one {\it extreme} EoS. 

The conclusion drawn from Fig. \ref{fig:yp_distr} is similar to the $v_s^2(n)$ results (see Fig. \ref{fig:vs_distr}):
it is evident that dataset 1 exhibits the narrowest prediction uncertainty, whereas dataset 2 demonstrates a considerable increase in uncertainty; the prediction uncertainties are similar for datasets 3 and 4. Definite conclusions require some statistics over a set of EoS, and, for that purpose, we are going to analyze the whole test set once again.\\

\begin{figure*}[!ht]		
\includegraphics[width=.33\linewidth]{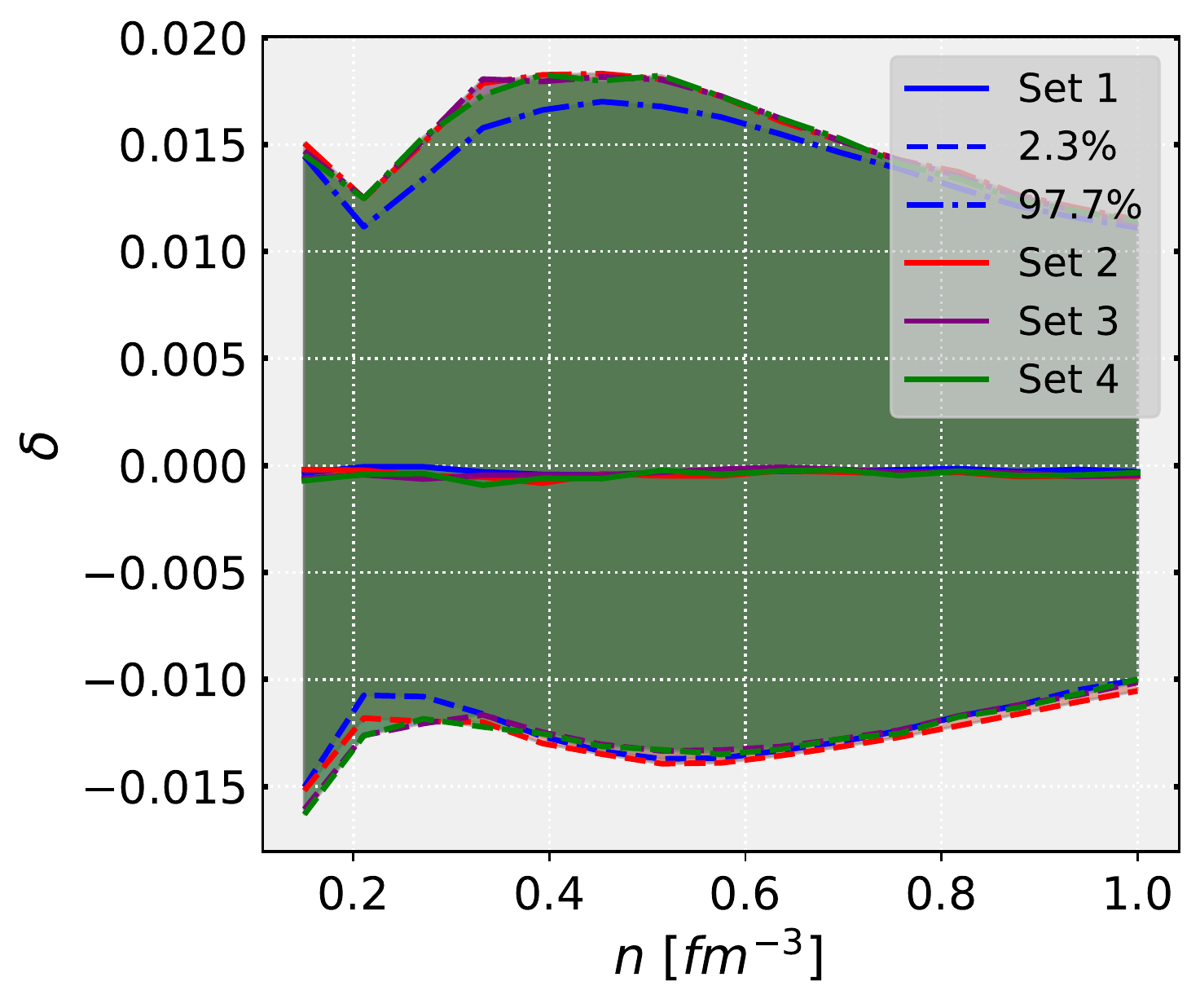} 
\includegraphics[width=.33\linewidth]{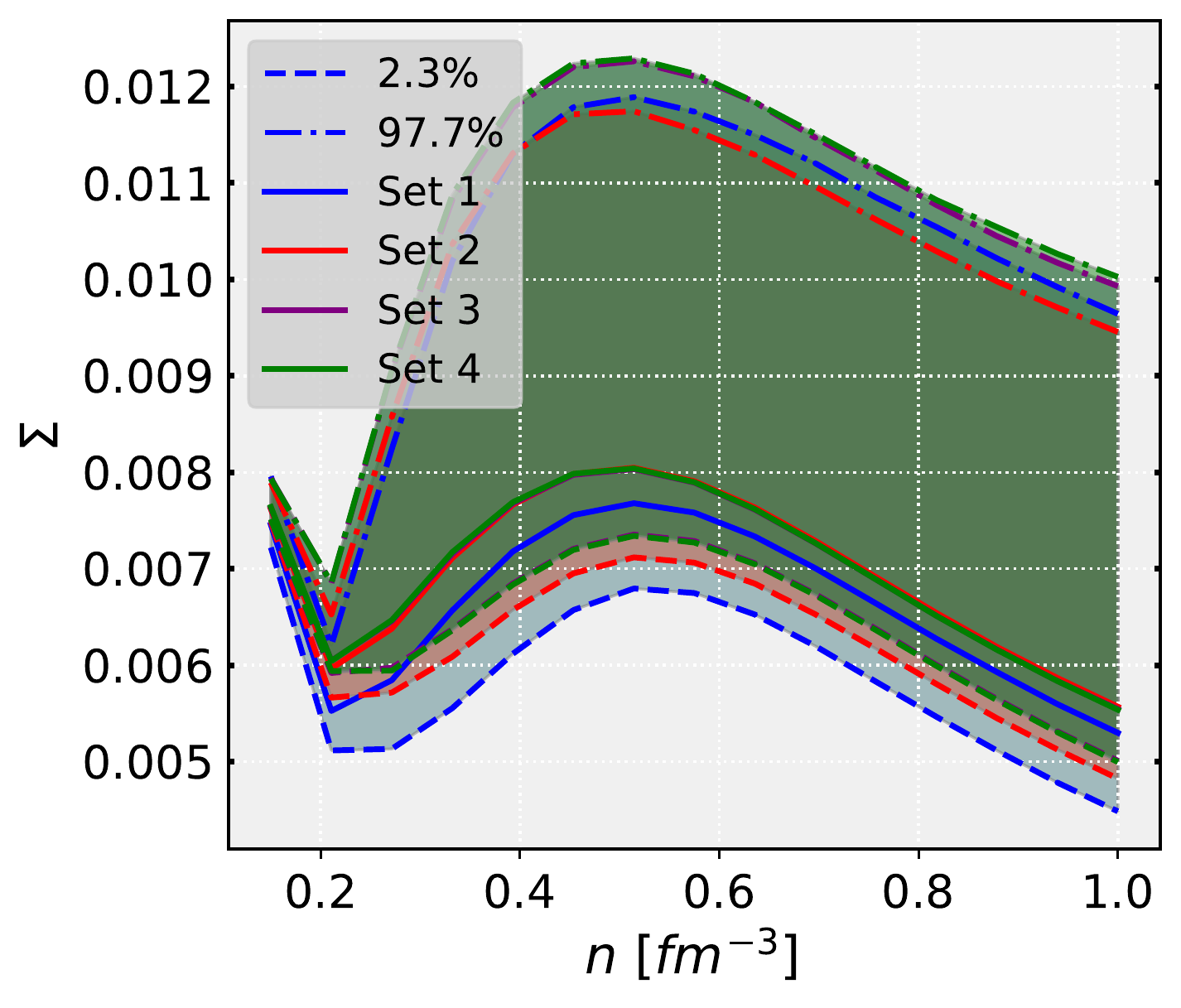}  
\includegraphics[width=.32\linewidth]{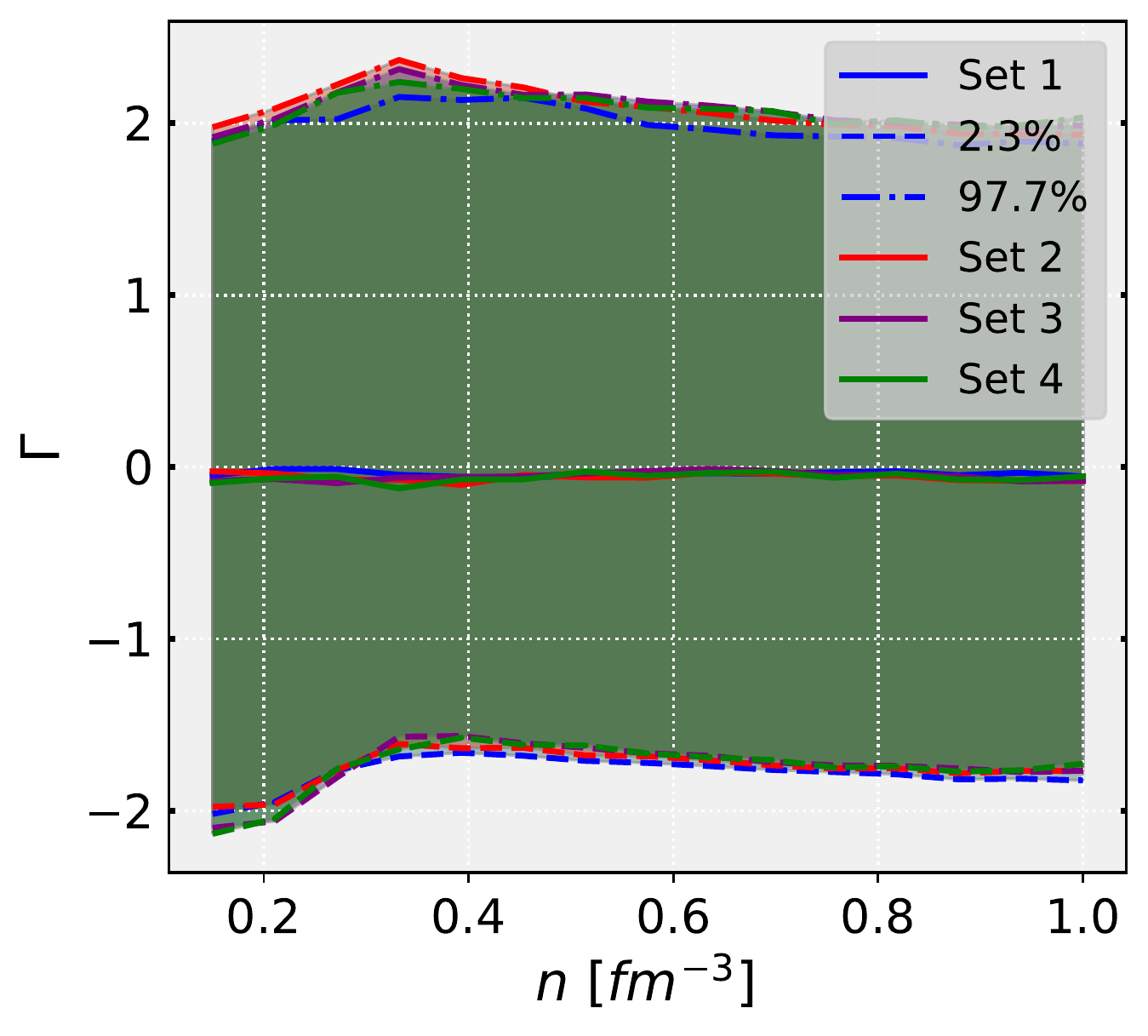}  
\caption{Median (solid line) and the 95.4\% confidence interval (dashed and dotted lines) for $\delta(n_k)=y_p(n_k)-y_p(n_k)^{\text{true}}$ (left), $\Sigma(n_k)=\sigma(n_k)$ (center), and $\Gamma(n_k)=\pc{y_p(n_k)-y_p(n_k)^{\text{true}}}/\sigma(n_k)$ (right). }
\label{fig:yp_sd}
\end{figure*}

\begin{figure}[!ht]		
\includegraphics[width=.95\linewidth]{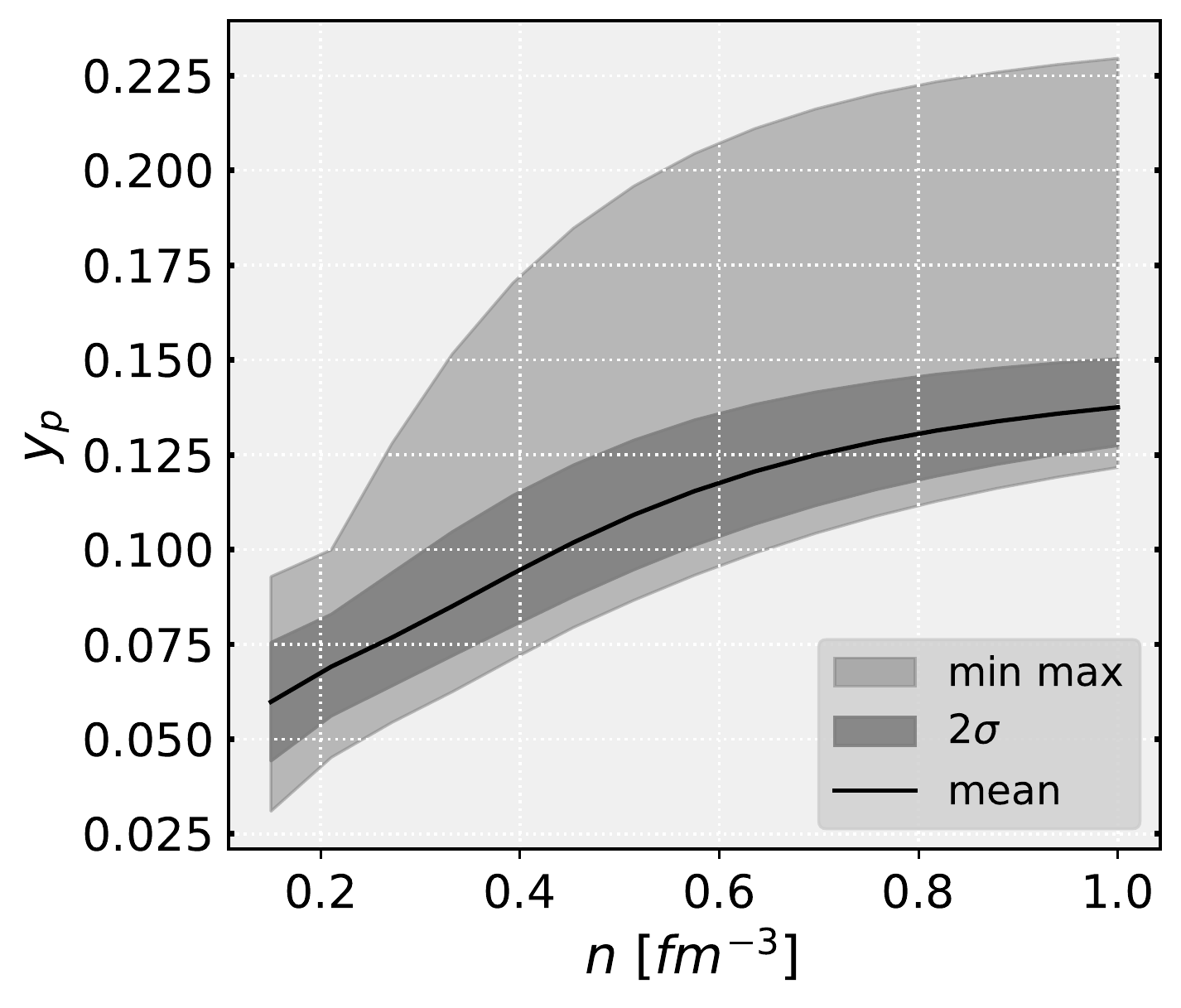} 
\caption{Some statistics of $y_p(n)$ calculated from the train dataset.}
\label{fig:yp_data_distr}
\end{figure}

Figure \ref{fig:yp_sd} shows the following quantities over the four datasets: the model residuals $\delta(n_k)=y_p(n_k)-y_p(n_k)^{\text{true}}$ (left), the standard deviation
$\Sigma(n_k)=\sigma(n_k)$ (center), and the normalized model residuals $\Gamma(n_k)=\pc{y_p(n_k)-y_p(n_k)^{\text{true}}}/\sigma(n_k)$ (right).
We calculate these quantities, at each density $n_k$, for each of the 2529 EoS from the test set. We see that the model has a broader residual spread around 0.4-0.5 fm$^{-3}$ that is counterbalanced by larger standard deviation values in this region - this is a direct conclusion when looking at the normalized model residuals steadiness. In other words, the model correctly captures and models the data statistics:
the BNN models capture larger prediction uncertainties in regions where $y_p(n)$ has a larger dispersion, as expected. The overall quality of the models prediction is seen in the density independence of the $\Gamma(n)$ (right panel) statistics, the models residuals are $95.4\%$ of the times between $2\sigma$. The insight of the train set statistics, in Fig. \ref{fig:yp_data_distr}, entails interpretability for $\delta(n)$ and particularly $\Sigma(n)$ behaviour, since it becomes clearer that $\sigma(n)$ of the train set has a non-monotonic behaviour, reaching a maximum value around 0.5 fm$^{-3}$ decreasing for lower/larger densities. The coverage probability for the $y_p(n)$ models are similar to the ones obtained for $v_s^2(n)$ (see Fig. \ref{fig:vs_cover}) and thus  the models are correctly estimating the distribution of the test set data.

\begin{figure}[!ht]		
\includegraphics[width=.95\linewidth]{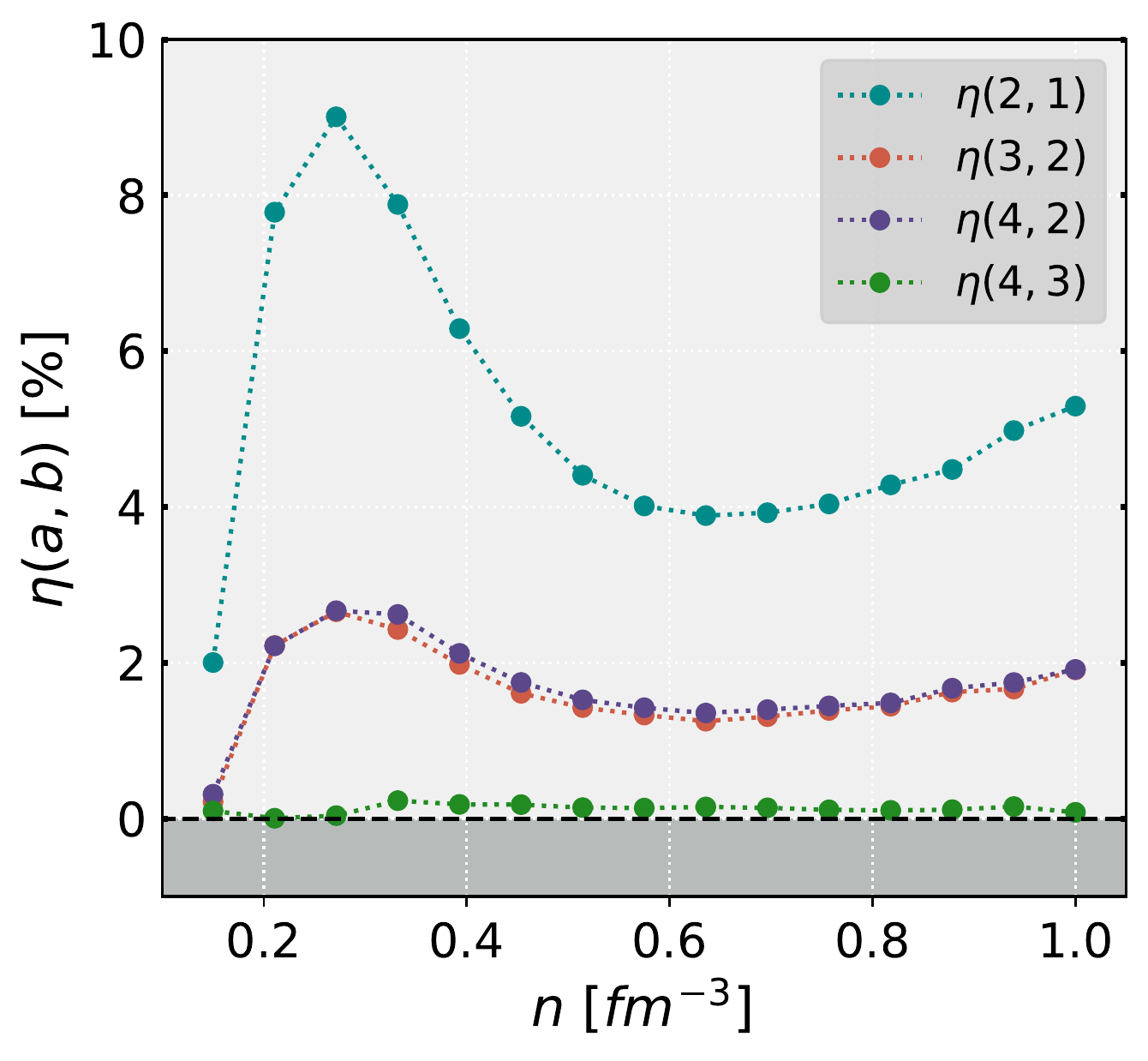}
\caption{Prediction uncertainty deviation $\eta\pr{a,b}$ between the $y_p(n)$ BNN models $a$ and $b$ (see text for details). }
\label{fig:yp_eta}
\end{figure}

The dataset comparison for $\eta\pr{a,b}(n)$ (see Eq. \ref{eq:eta}) is displayed in Fig. \ref{fig:yp_eta}.
Firstly, we observe that $\eta\pr{2,1}$ exhibits a behavior similar to $v_s^2$. However, it reaches its maximum value earlier at $n_3$. Interestingly, this is precisely where $R(M)$ and $y_p(n)$ demonstrate the highest correlation, as explained in Annex \ref{append_1}. 
When comparing $\eta[2,1]$ with $\eta[4,3]$,  the behaviour is consistent with the one observed in the speed of sound, albeit this time is even more outstanding, which comes back again to the correlation, that is almost zero for $\Lambda(M)$ and $y_p(n)$. Based on this observation, we can speculate that $\Lambda$ does not contribute with significant information to the model, especially since this time the ratio between $\eta[4,3]$ and $\eta[2,1]$ deviates even further from the expected one-quarter proportion, that comes from the change in the input vector, than it did on the $v_s^2$. Increasing the model's complexity without adding substantial information can lead to increased confusion and greater uncertainty, as evident in $\eta[3,2]$ and $\eta[4,2]$. These two cases exhibit positive values of $\eta$, indicating higher uncertainty for sets 3 and 4 when compared with set 2.\\

\subsection{BNNs epistemic and aleatoric uncertainties}
Let us briefly analyze how the prediction uncertainty of BNNs is modelled and its different components. We are going to focus on the $y_p(n)$ BNN models. However, all the features discussed below are also seen for the $v_s^2(n)$ models. We have seen in Eq. \ref{eq:errors} of Sec. \ref{bnn} that the prediction variance $\hat{\bm{\sigma}}^2$ is a combination of two terms,
\begin{equation*}
\hat{\bm{\sigma}}^2=\hat{\bm{\sigma}}^2_{\text{alea}}+ \hat{\bm{\sigma}}^2_{\text{epist}},
     \label{eq:error}
\end{equation*}
where the aleatoric uncertainty $\hat{\bm{\sigma}}^2_{\text{alea}}$ measures the mean variance of the models' ensemble, while the epistemic uncertainty $\hat{\bm{\sigma}}^2_{\text{epist}}$ measures the spread of the models around the ensemble mean $\hat{\bm{\mu}}$. 
The epistemic uncertainty arises from limited information or data and is encoded on the posterior probability $P(\bm{\theta}|D)$, i.e., the model distribution.
On the other hand, aleatoric uncertainty 
is due to the inherent randomness of the dataset
and is encoded into the data likelihood $P(\bm{y^*}|\bm{x^*},\bm{\theta})$.
While the epistemic uncertainty decreases when more data is available, the aleatoric uncertainty value does not depend on the amount of data has it is a property of the generating data process.\\

\begin{figure}[!ht]		
\includegraphics[width=1\linewidth]{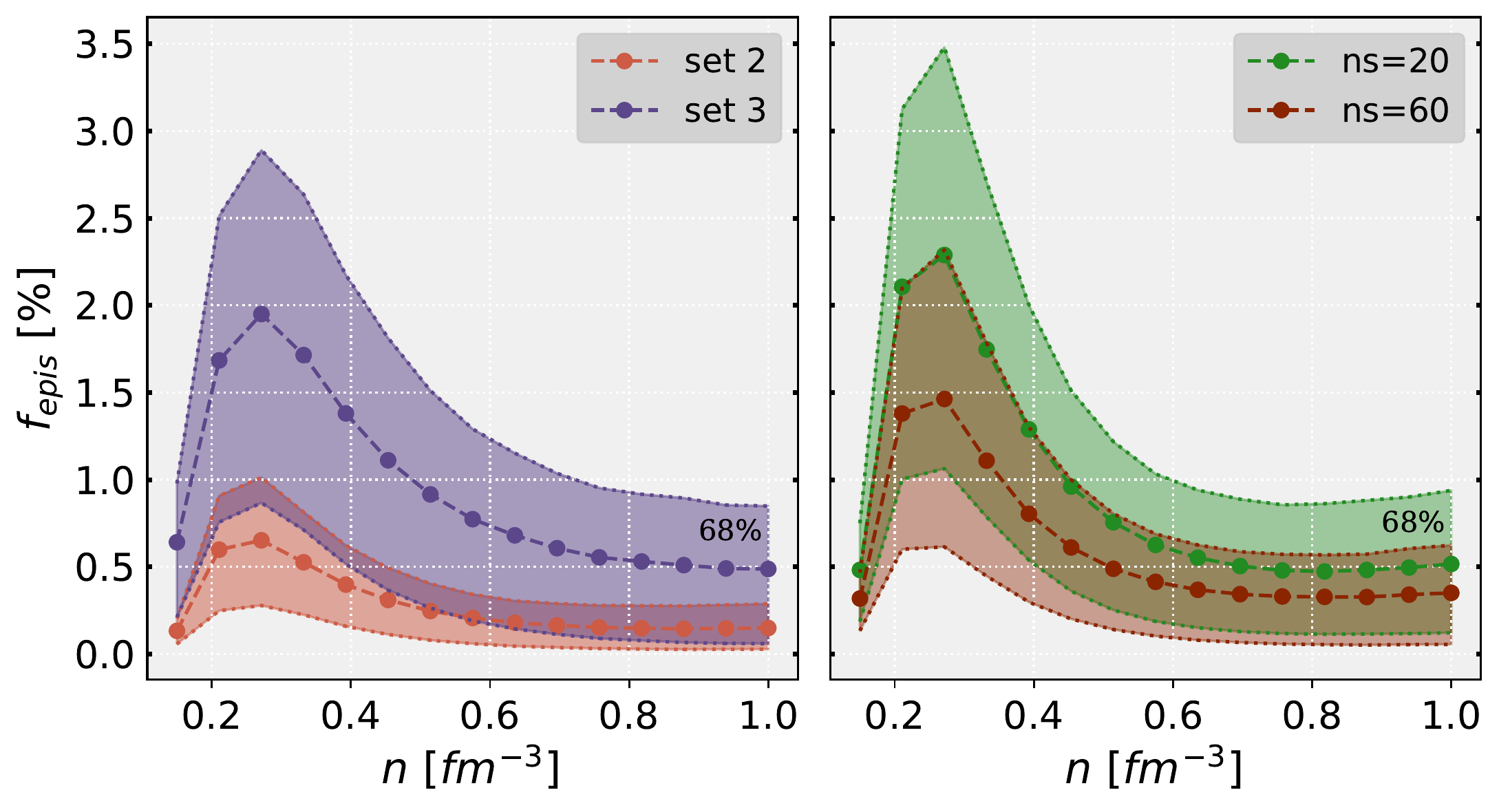}
\caption{The pdf of $f_{\text{epist}}=(\hat{\bm{\sigma}}^2_{\text{epist}}/\hat{\bm{\sigma}}^2)\times 100\%$ for $y_p(n)$ BNN models in sets 2 and 3 (left) and for set 1 model but trained in different training datasets with different number of mock observations $n_s$ (right).}
\label{fig:erros}
\end{figure}

To analyze the proportions of both uncertainty types on the total prediction variance, we calculate the epistemic percentage as $f_{\text{epist}}=(\hat{\bm{\sigma}}^2_{\text{epist}}/\hat{\bm{\sigma}}^2)\times 100\%$ .
Using the BNNs models for $y_p(n)$, we show in the left panel of Fig. \ref{fig:erros} the mean (dashed lines) and 68\% confidence interval region (colored regions) of  $f_{\text{epist}}$ across the entire test (2 and 3) sets.  Additionally, 
the right panel shows $f_{\text{epist}}$ using BNN models trained on set 1 but with different numbers of mock observations: $n_s=20$ and $n_s=60$ (the value used in this work).
The following conclusions can be drawn: i) the prediction variance $\hat{\bm{\sigma}}^2$ is composed mainly of aleatoric uncertainty (left panel), around 95\%, and it is due to the already high number of mock observations $n_s=60$; ii) the right panel shows that decreasing the number of mock observations $n_s$, and thus the total number of training points, increases the epistemic uncertainty. The epistemic uncertainty converges to zero when the data points go to infinity;
iii) left panel also shows that  $f_{\text{epist}}$ is smaller for set 2 (orange) than set 3 (purple) because the input dimensions increase from 10 to 20, which is reflected on the posterior $P(\bm{\theta}|D)$.
Lastly, let us argue why the epistemic uncertainty is larger at densities 0.2 -- 0.4 fm $^{-3}$. When constructing the predicting ensemble,  $P(\bm{y^*}|\bm{x^*}, D)=\frac{1}{N} \sum_{n=1}^{N} P(\bm{y^*}|\bm{x^*},\bm{\theta}^{(n)})$, by sampling from the variational posterior, $ \bm{\theta}^{(n)} \sim q_\phi(\bm{\theta})$, the density points $n_k$ with larger correlation with $y_p(n)$ are much more sensitive to model sampling than other density points where correlations are much weaker.

\section{Prediction for the DD2 nuclear model}
\begin{figure}[!ht]		
\includegraphics[width=.91\linewidth]{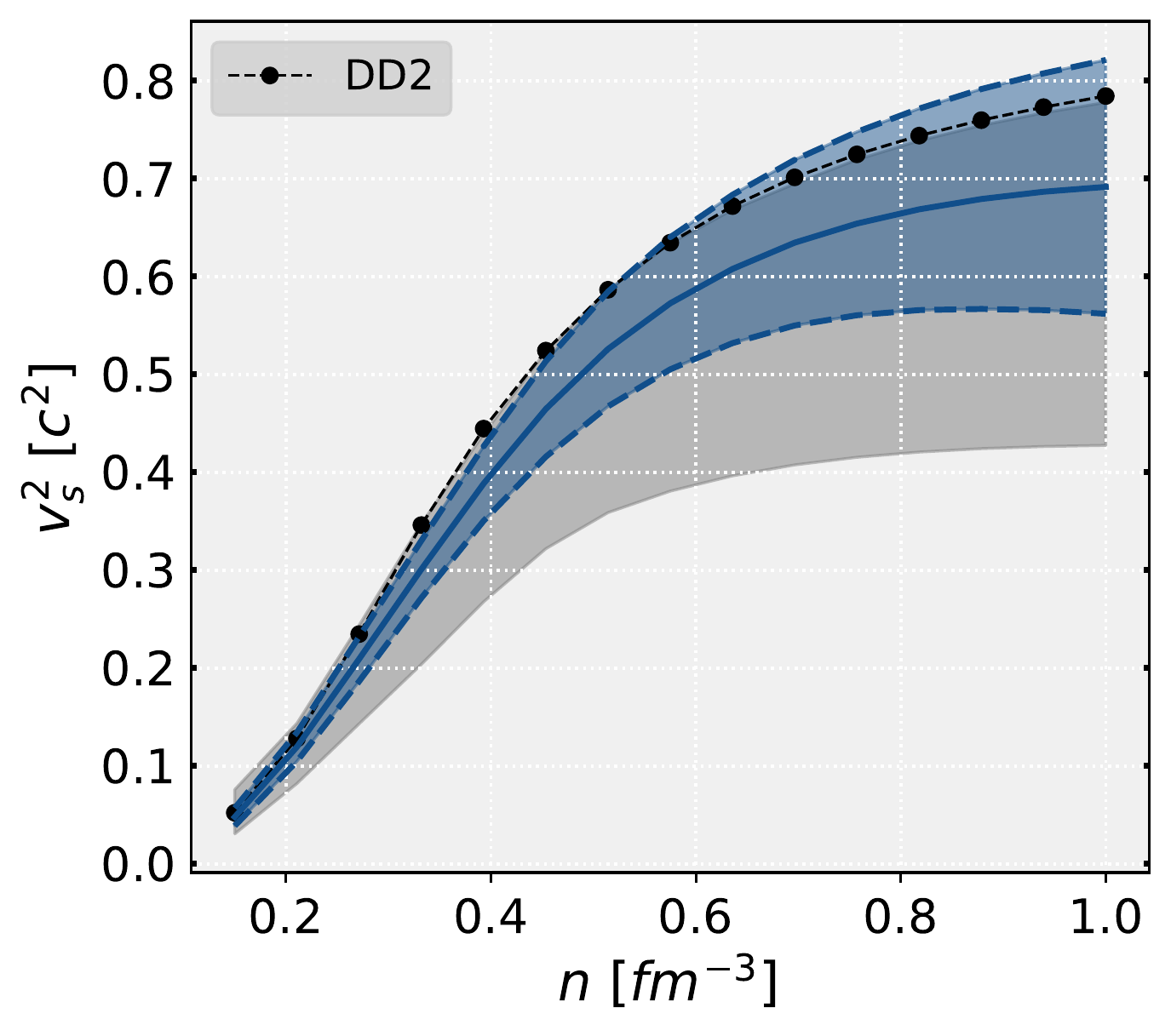}\\
\includegraphics[width=.95\linewidth]{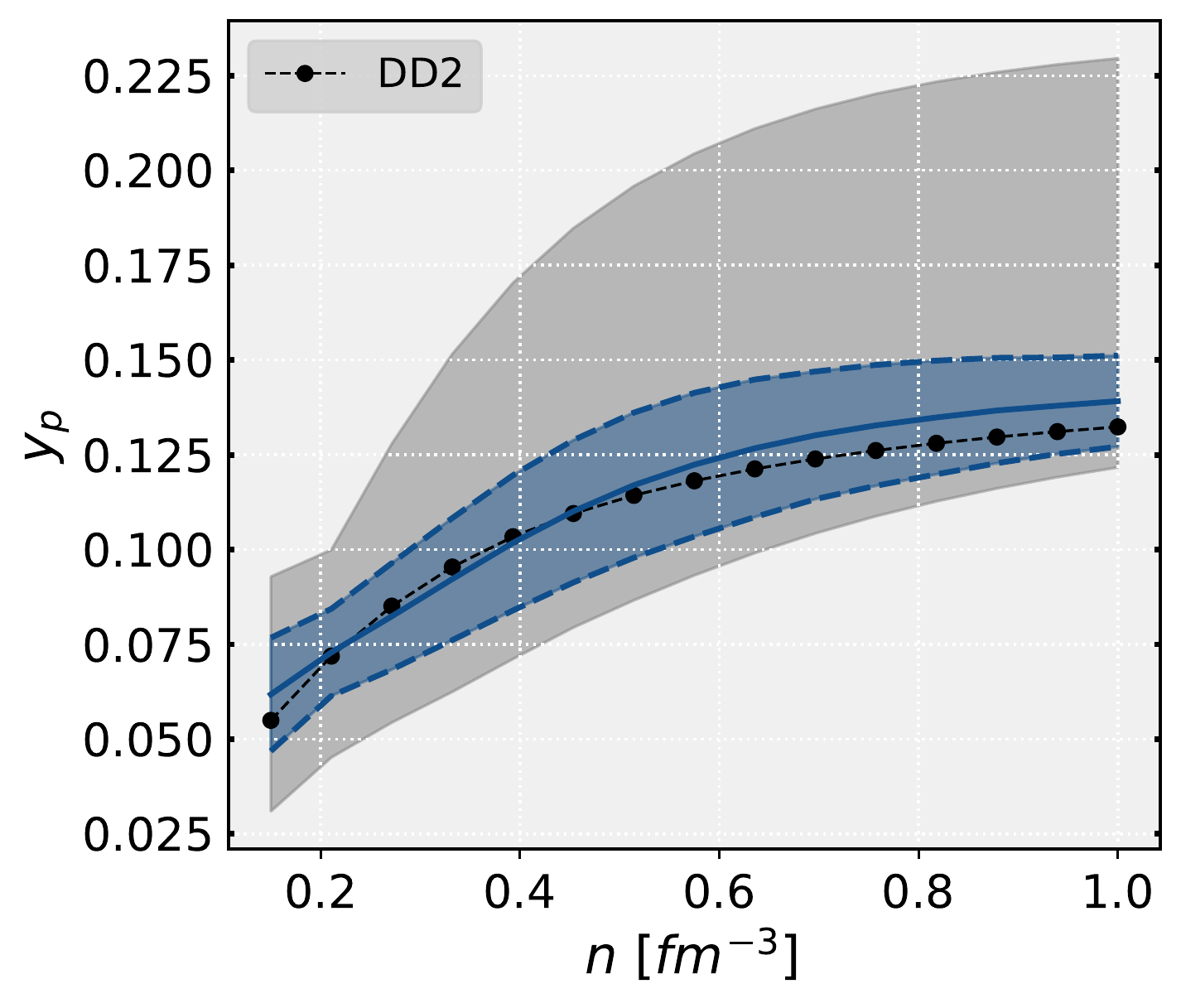}
\caption{The BNN model predictions, $v_s^2$ (upper) and $y_p$ (lower), for one mock observation ($n_s=1$) of the DD2 EoS, the blue area represents the {95.4\%} confidence interval, and the solid line the mean.} 
\label{fig:dd2}
\end{figure}

As a final test, we applied the BNN model (trained on set 1, see Tab. \ref{tab:sets}) to a  nuclear model with different properties from the ones used to train, in particular, obtained within a different microscopic description of nuclear matter.
We select the DD2 model which is a generalized relativistic mean-field model (RMF) model with density-dependent couplings \cite{Typel:2009sy}, which has been calibrated to describe properties of finite nuclei.
One key difference between the DD2 and the RMF family we used to generate the set of EoS consists of the high density behavior of the symmetry energy. In DD2 model, the coupling to the $\rho$-meson that defines the isovector channel of the EoS goes to zero at sufficiently high densities, favoring very asymmetric matter. One of the main consequences is that nucleonic direct Urca processes inside NS are not predicted by DD2 \cite{Fortin:2016hny,Fortin:2021umb}. 
Another noticeable difference between the DD2 class of models   and the class of models used to train the BNN is the behavior of the speed of sound with density:  for DD2 like models the speed of sound increases monotonically, although it remains always well below $c$, while  for the class of models used to train BNN the speed of sound flattens or even decreases above $\sim 3\rho_0$. These two differences will be reflected in the performance of the BNN model.\\\\
After selecting the DD2 EoS, following the statistical procedure described in Sec. \ref{dataset}, we generated one mock observation ($n_s=1$) using the dataset 1 properties, which is the one with lower $\sigma_R$ and does not contain the information about $\Lambda$. The BNN model predictions for the speed of sound (top panel) and proton fraction (lower panel) are shown in Fig. \ref{fig:dd2}. Despite the DD2 lying outside the training values (grey region) for the speed of sound, the model prediction uncertainty extends beyond the training maximum values and almost contains completely the DD2 results. The $y_p(n)$ prediction is quite good, being the prediction of the mean value close to the real one. The DD2 proton fraction reflects the property described above concerning its favoring large neutron-proton asymmetries due to the behavior of the isovector channel. However, our BNN model was able to capture this behavior.
Despite the results being quite compelling, there are some crucial points that we would like to point out. During the above test stage, we generated just one mock observation ($n_s=1$) from the DD2 $M(R)$ curve to simulate a real case scenario, where a very limited number of NS observations are accessible. Being the sampling procedure of generating one mock observation ($n_s=1$), i.e., five $M_i(R_i)$ values, a random process, implies that different samples will originate different predictions ($v_s^2(n)$ is much more sensitive than $y_p(n)$, since the DD2 target is completely inside the training values region). This is a somehow expected behaviour since we are trying to characterize the whole $M(R)$ with only 5 random $(M,R)$ values. While a given sample may be sufficient to inform the general dependence of the $M(R)$ (like the one we generated), others might be as well almost uninformative of the actual $M(R)$ curve, i.e., a sample where all five mock observations $M_i(R_i)$ cluster around the same $M$ value. This is a general problem that shows up regardless of the inference model or framework: inferring the EoS from a very limited number of NS observations. The BNN performance assessment for the DD2 EoS would
get more reliable as the number of points $M_i(R_i)$, which compose each mock observation (5 in the present work), increase since a random sample, in that case, would be much more informative of the true $M(R)$ curve.

\section{Conclusions \label{conclusions}}
We have explored Bayesian Neural Networks (BNNs), which is a probabilistic machine learning model, to predict the proton fraction and speed of sound of neutron star matter from a set of NS mock observations. This method is based upon the usual neural networks but with the crucial advantage of attributing an uncertainty measurement to its predictions. Our EoS dataset was generated from a relativistic mean field approach through a 
Bayesian framework, where constraints from nuclear matter properties and NS observations were applied.  The choice of a specific microscopic nuclear model, instead of a more flexible EoS parameterization, as the ones discussed for instance in \cite{Annala2019}, is justified because we want to analyze the possibility of inferring the neutron star composition, specifically, the proton fraction, from NS observations. From the set of 25287 EoS, four different mock observational sets, simulating four different scenarios of mock observational uncertainties, were generated. Two of them are only composed of $M(R)$ simulated observations and the other two  have also information regarding $\Lambda(M)$. In the end, 8 different BNNs were trained to predict the $v_s^2(n)$ and $y_p(n)$ in each of the 4 datasets. 

With this study, we have shown that using BNNs, the measurements of the mass and radius of five neutron stars allow us to recover information from the equation of state of nuclear matter with associated uncertainty, not only for a quantity that is more connected with the isoscalar behavior of the EoS, the speed of sound, but also for the proton fraction, a property that is determined by the isovector behavior of the EOS. In several recent works, the attempt to determine the proton fraction from the mass radius measurements was unsuccessful  \cite{Tovar2021,Imam2021,Mondal2021}. In all these descriptions a polynomial expansion of the EoS until the third or fourth order has been considered. In \cite{Mondal2021},  this was attributed to the existence of multiple solutions. The authors of \cite{Imam2021} identify
the correlations among higher-order parameters as a difficulty. The BNN approach allows the model to learn the full density dependence of the EoS avoiding the short comes of the density expansion with a finite number of terms.
It was shown that the uncertainty associated with the predicted quantities is particularly sensitive to the precision of the observational data if some kind of correlation exists between the data and the property that is being calculated. For the speed of sound this was reflected in a larger sensitivity for densities below three times saturation density, where the NS radius is strongly correlated with the speed of sound as discussed in \cite{Ferreira:2019bgy}. It was also shown that adding extra observational mock data, in particular, the tidal deformability, could decrease the uncertainty associated with the prediction, but not always. There was a clear improvement for the speed of sound but not for the proton fraction. Too scattered data does not bring an improvement on the uncertainty determination due to the increase of the complexity of the model compared with the quality of the data. It is important to point out that the improvement attained with the tidal deformability  data with a smaller uncertainty worked for the speed of sound  because it shows a correlation with the tidal deformability for densities of the order of twice saturation density similar to the one with the radius.  This correlation does not exist between the proton fraction and the tidal deformability so no improvement in the proton fraction prediction was attained when introducing the tidal deformability observation. The
proton fraction has shown some sensitivity at twice saturation density to the radius uncertainty and this can be traced back to the existing correlation of low mass star radius with the symmetry energy slope \cite{Alam:2016cli}, a quantity that strongly determines the proton fraction. This correlation weakens quickly with the increase of the NS mass, and is much weaker with the tidal deformability. We have also tested the BNN model with a  mock measurement obtained from the DD2 EoS generated with a microscopic framework different from the one used to generate the EoS used to train the BNN model. The results have confirmed the validity of the model and its predicting power.

We have been very conservative concerning the uncertainties attached to the observations. In the future, observatories such as STROBE-X  \cite{STROBE-X}  and eXTP \cite{eXTP} may give us  radius measurements with uncertainties as small as 2\%-5\% and this will improve the predictions as demonstrated in the present study.

There are several potential paths for further improvement and exploration in this work. One possibility is to extend the analysis to include other properties of neutron stars and investigate their relationship with observable quantities. Another possibility for improvement as discussed in the results obtained for the DD2 model, is increasing the number of observable pairs used as input can enhance the model's performance.  However, it is worth noting that in the case of Bayesian neural networks (BNNs), expanding the number of pairs introduces a greater increase in the model parameters compared to traditional architectures, which is why we used a lesser amount of pairs compared with previous articles using conventional neural networks, as demonstrated in studies like \cite{fujimoto2021extensive} and related articles. Furthermore, for the stochastic model, it would be interesting to improve the prior as mentioned in section \ref{sec:vi}. 

\section*{ACKNOWLEDGMENTS} 
This work was partially supported by national funds from FCT (Fundação para a Ciência e a Tecnologia, I.P, Portugal) under Projects No. UIDP/\-04564/\-2020, No. UIDB/\-04564/\-2020 and 2022.06460.PTDC. 

\onecolumngrid
\appendix
    
\section{Correlation between NS properties and EoS}
\begin{figure}[!ht]
    \includegraphics[width=0.45\textwidth]{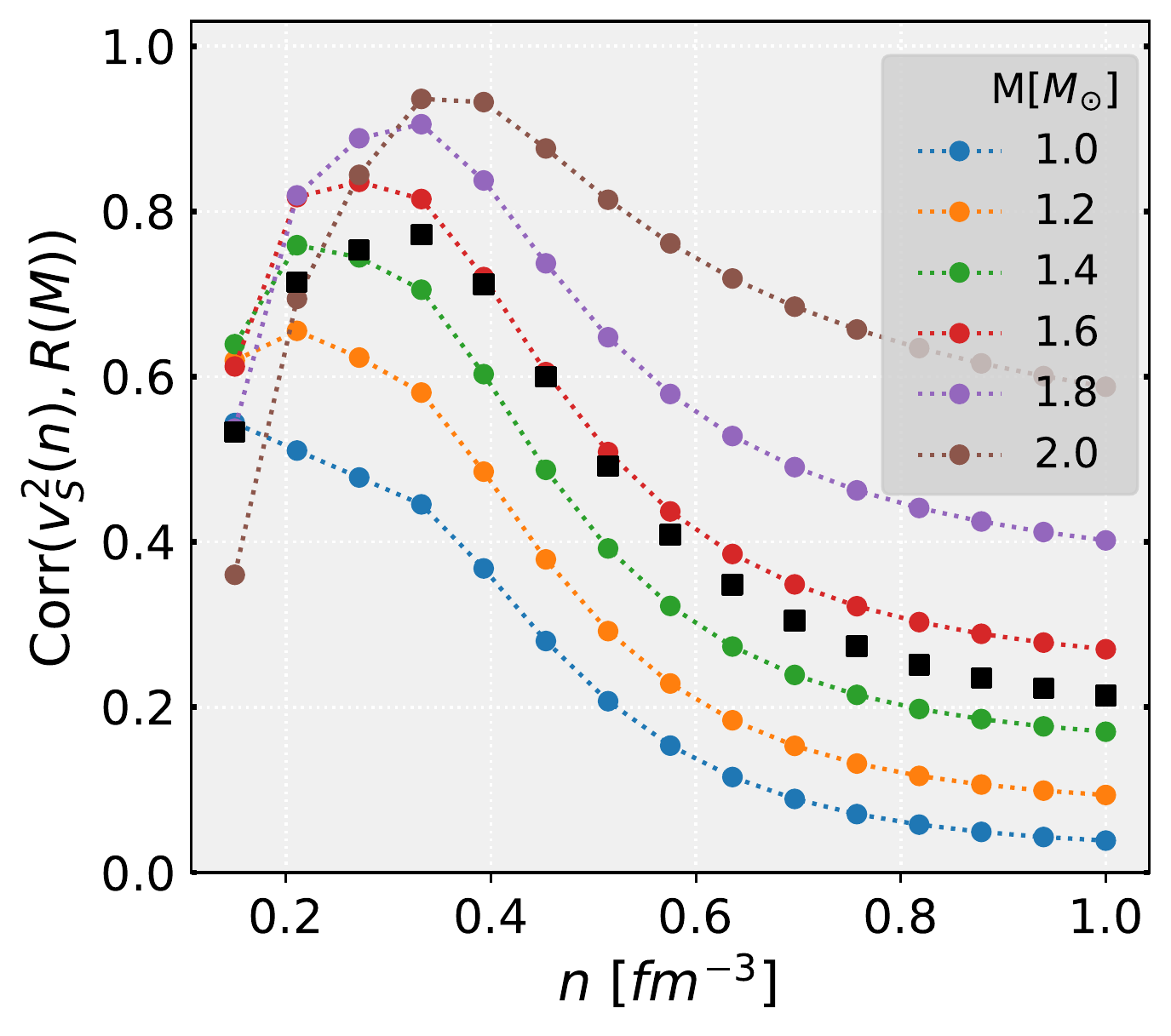}
\includegraphics[width=0.45\textwidth]{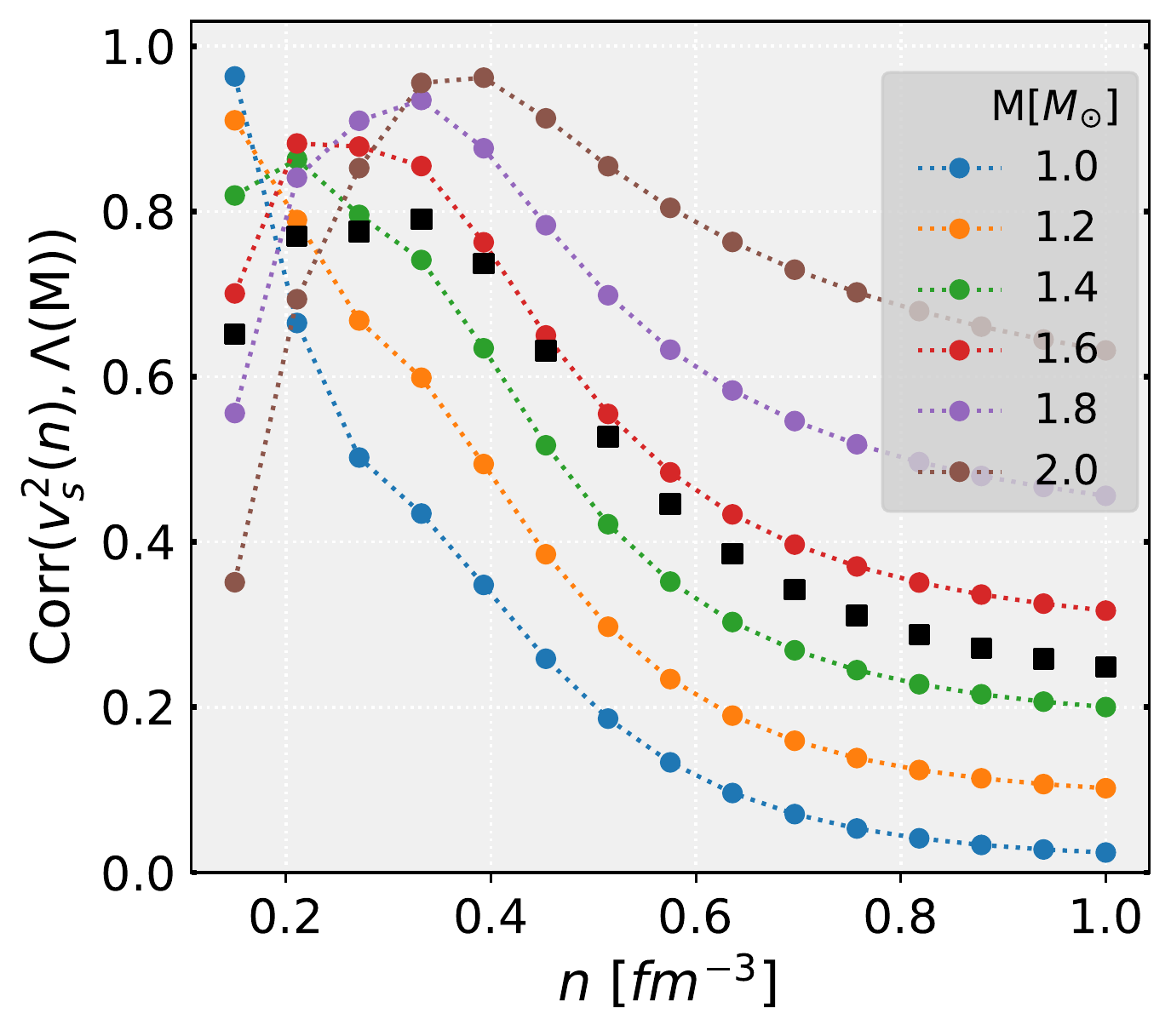}
    \caption{Correlation between $v_s^2(n)$ and $R(M)$ (left) or $\Lambda(M)$ (right) for fixed mass values. In black squares we show the mean correlation value for $M/M_{\odot}\in [1,2.2]$.}
    \label{fig:corr_vs}
\end{figure}

\begin{figure}[!ht]
    \includegraphics[width=0.45\textwidth]{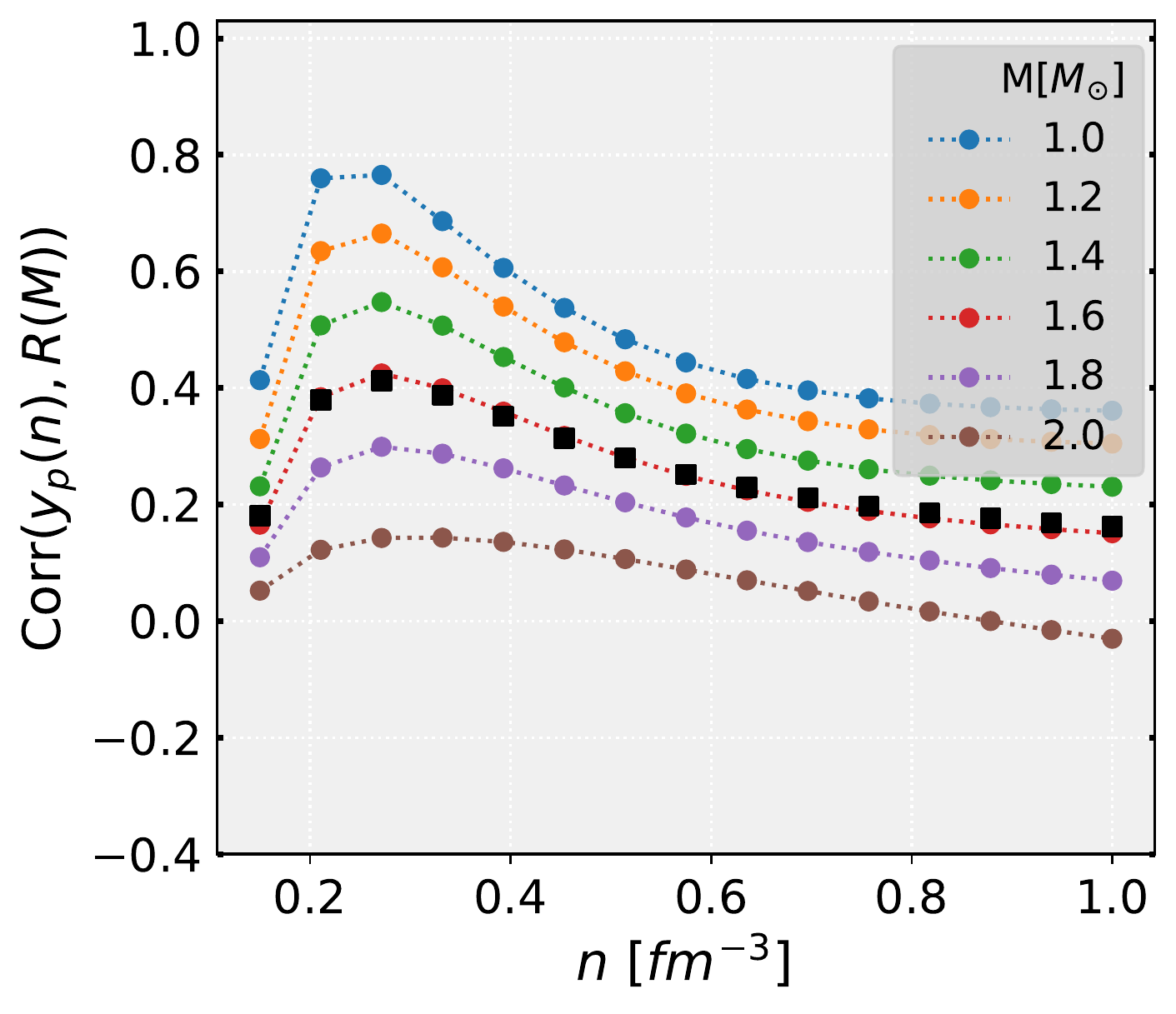}
    \includegraphics[width=0.45\textwidth]{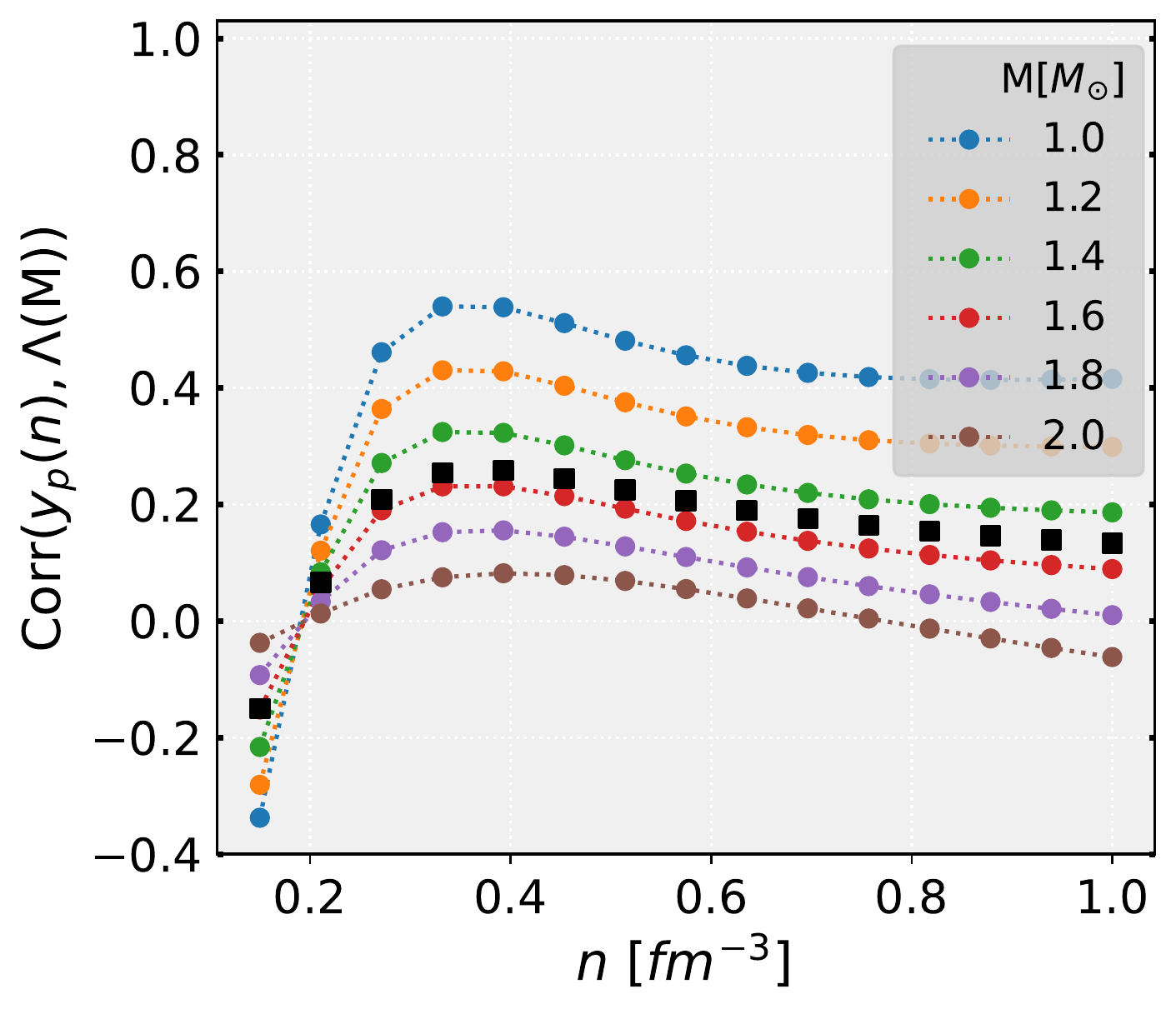}
    \caption{Correlation between $y_p$ and $R(M)$ (left) or $\Lambda(M)$ (right) for fixed mass values. 
    In black squares we show the mean correlation value for $M/M_{\odot}\in [1,2.2]$.}
    \label{fig:corr_yp}
\end{figure}

Figure \ref{fig:corr_vs} shows the Pearson correlation coefficient between $v_s^2(n)$ and $R(M)$ (left panel) and $\Lambda(M)$ (right panel) for specific NS masses (colors) and the average value, by considering  $M/M_{\odot}\in [1,2.2]$. The same is performed for the proton fraction in Fig. \ref{fig:corr_yp}. The Pearson correlation is calculated as $\text{Corr}(a,b)=\text{Cov}(a,b)/(\sigma_{a} \sigma_{b})$, where $a$ consists of $v_s^2$ and $y_p$ and $b$ of $R$ and $\Lambda$. Note, however, that this correlation measure is only sensitive to linear dependencies, and higher order ones can be missed.  These correlations have been discussed in \cite{Ferreira:2019bgy}.
\label{append_1}

\twocolumngrid
\bibliographystyle{apsrev4-1}
%

\end{document}